%% file: 0.head.tex
\documentclass[10pt,journal,compsoc]{IEEEtran}
\usepackage{cite}
\usepackage{amsmath,amssymb,amsfonts}
\usepackage{algorithmic}
\usepackage{graphicx}
\usepackage[justification=centering]{caption}
\usepackage{xcolor}

\usepackage{booktabs} 
\usepackage{subfigure}
\usepackage{amsmath}
\usepackage[normalem]{ulem}
\usepackage{enumitem}
\usepackage{multirow}
\usepackage[nomargin,inline,marginclue,draft]{fixme}
\usepackage{balance}
\usepackage{changepage}

\newlength\savedwidth
\newcommand\whline{\noalign{\global\savedwidth\arrayrulewidth
		\global\arrayrulewidth 1.1pt}%
	\hline
	\noalign{\global\arrayrulewidth\savedwidth}}
\newcommand{\para}[1]{{\vspace{4pt} \bf \noindent #1 \hspace{10pt}}}

\def\BibTeX{{\rm B\kern-.05em{\sc i\kern-.025em b}\kern-.08em
    T\kern-.1667em\lower.7ex\hbox{E}\kern-.125emX}}
\begin{document}

\title{Learning to Recommend with Multiple Cascading Behaviors}

\author{Chen~Gao,
        Xiangnan~He,
        Dahua~Gan,
        Xiangning~Chen,
        Fuli~Feng,
        Yong~Li,~\IEEEmembership{Senior~Member,~IEEE,}
        Tat-Seng~Chua,
        Lina~Yao,
        Yang~Song,
        and~Depeng~Jin,~\IEEEmembership{Member,~IEEE}
\IEEEcompsocitemizethanks{\IEEEcompsocthanksitem C. Gao, X. Chen, Y. Li, and D. Jin are with Beijing National Research Center for Information Science and Technology, Department of Electronic Engineering, Tsinghua University, Beijing 100084, China.
E-mail: \{gc16, cxn15\}@mails.tsinghua.edu.cn,\{liyong07,jindp\}@tsinghua.edu.cn
\protect\\
\IEEEcompsocthanksitem X. He is with School of Information Science and Technology, University of Science and Technology of China, Hefei, China.
E-mail: xiangnanhe@gmail.com.
\protect\\ 
\IEEEcompsocthanksitem D. Gan is with the School of Computer Science, Carnegie Mellon University, USA. E-mail:dgan@andrew.cmu.edu.
\protect\\
\IEEEcompsocthanksitem F. Feng and TS. Chua are with School of Computing, National University of Singapore, Computing 1, Computing Drive, 117417, Singapore.
E-mail: fulifeng93@gmail.com, dcscts@nus.edu.sg.
\protect\\
\IEEEcompsocthanksitem L. Yao and Y. Song are with the School of Computer Science and Engineering, the University of New South Wales (UNSW), Australia.
E-mail: \{lina.yao,yang.song1\}@unsw.edu.au.
}
}

\markboth{IEEE TRANSACTIONS ON KNOWLEDGE AND DATA ENGINEERING}
{Shell \MakeLowercase{\textit{et al.}}: Bare Demo of IEEEtran.cls for Computer Society Journals}
\IEEEtitleabstractindextext{%
\begin{abstract}
Most existing recommender systems leverage user behavior data of one type only, such as the purchase behavior in E-commerce that is directly related to the business KPI (Key Performance Indicator) of conversion rate. Besides the key behavioral data, we argue that other forms of user behaviors also provide valuable signal, such as views, clicks, adding a product to shopping carts and so on. They should be taken into account properly to provide quality recommendation for users. 

In this work, we contribute a new solution named NMTR (short for \textbf{N}eural \textbf{M}ulti-\textbf{T}ask \textbf{R}ecommendation) for learning recommender systems from user multi-behavior data. We develop a neural network model to capture the complicated and multi-type interactions between users and items. In particular, our model accounts for the cascading relationship among  different types of behaviors (e.g., a user must click on a product before purchasing it). To fully exploit the signal in the data of multiple types of behaviors, we perform a joint optimization based on the multi-task learning framework, where the optimization on a behavior is treated as a task. Extensive experiments on two real-world datasets demonstrate that NMTR significantly outperforms state-of-the-art recommender systems that are designed to learn from both single-behavior data and multi-behavior data. Further analysis shows that modeling multiple behaviors is particularly useful for providing recommendation for sparse users that have very few interactions. 

\end{abstract}

}

\maketitle
\IEEEdisplaynontitleabstractindextext
\IEEEpeerreviewmaketitle

\input{1.introduction}
\input{2.motivation}

\input{3.method}

\input{4.evaluation}
\input{5.relatedwork}

\vspace{-0.20cm}
\section{Conclusion and future work}\label{sec:conclusion}
In this work, we designed a recommendation system to exploit multiple types of user behaviors. We proposed a neural network method named NMTR, which combines the recent advances of NCF modeling and the efficacy of multi-task learning. We conducted extensive experiments on two real-world datasets and demonstrated the effectiveness of our NMTR method on multiple recommender models. 
This work makes the first step towards understanding how to integrate the rich semantics of users' multiple behaviors into recommender systems. With increasing kinds of user behaviors on the Web, we believe multi-behavior recommendation is an important topic and will attract more attention in the future. 

As for future work, we will perform online evaluation of our NMTR method through A/B tests, and focus more on the practical issues of online learning and incremental learning. 
On the other hand, we will study multi-behavior recommendation in the scenarios that user behaviors cannot form a full-order cascading relation. These behaviors not only contain the normal interactions between users and items, but may also include social interactions among users, such as sharing, following, etc. It is interesting to investigate how to integrate these heterogeneous kinds of user behaviors into a unified recommendation framework. 
Lastly, we will study time-aware models to capture the evolution of user preference in multi-behavior recommendation, especially for capturing dynamic user interests with RNN-based or other models for better recommendation.

\vspace{-0.50cm}
\section*{Acknowledgement}
	This work was supported in part by The National Key Research and Development Program of China under grant 2017YFE0112300, the National Nature Science Foundation of China under 61861136003, 61621091 and 61673237, Beijing National Research Center for Information Science and Technology under 20031887521, and research fund of Tsinghua University - Tencent Joint Laboratory for Internet Innovation Technology. This research is also part of NExT research which is supported by the National Research Foundation, Prime Minister’s Office, Singapore under its IRC@SG Funding Initiative. This work is also supported by the National Natural Science Foundation of China (61972372).

\bibliographystyle{IEEEtran}
\bibliography{bibliography}

\end{document}

%% file: 1.introduction.tex
\IEEEraisesectionheading{\section{Introduction}}\label{Sec:Intro}

\IEEEPARstart{I}{n} online information systems, users interact with a system in a variety of forms. For example, in an E-commerce website, a user can click on a product, add a product to shopping cart, purchase a product and so on. 

In traditional recommender systems, only user-item interaction data of one behavior type is considered for collaborative filtering, such as the purchase behavior in E-commerce and the rating behavior on movies~\cite{rendle2009bpr,he2017neural}. While it is particularly useful to optimize a recommender model on the data that is directly related to the business KPI, the other forms of behaviors should not be neglected, since they also provide valuable signal on a user's preference.

Existing approaches for multi-behavior recommendation can be divided into two categories. 
The first category is based on collective matrix factorization (CMF)~\cite{CMF,park2017also, krohn2012multi, zhao2015improving}, which extends the matrix factorization (MF) method to jointly factorize multiple behavior matrices.
In MF, a user (or an item) is described as an embedding vector to encode her preference (or its property), and a user-item interaction is estimated as the inner product of the user embedding and item embedding. 
To correlate MF on multiple behavior matrices, it is essential to share the embedding matrix of entities of one side (e.g., items), and let the entities of the other side (e.g., users) learn different embedding matrices for different types of behaviors.  

The second category approaches the problem from the perspective of learning~\cite{loni2016bayesian,qiu2018bprh}. To learn recommender models from the (implicit) data of interactions, it is natural to assume that a user's interacted items should be more preferable over the non-interacted items. Bayesian Personalized Ranking (BPR)~\cite{rendle2009bpr} is a representative method that implements the assumption of relative preference; it is then extended to address multi-behavior recommendation~\cite{loni2016bayesian} by enriching the training data of relative preference from the multi-behavior data.

Despite effectiveness, we argue that existing models for multi-behavior recommendation suffer from three limitations.
\begin{itemize}[leftmargin=*]
	\item \textbf{Lack of behavior semantics}. Each behavior type has its own semantics and contexts, and more importantly, there exist strong ordinal relations among different behavior types. For example, the behaviors may represent the action sequence of a user on a product: 
	click is not likely to happen after add-to-cart; add-to-cart behavior is not likely to happen after a purchase.
	 Moreover, the semantics make some intermediate feedback rather meaningful, such as the products that are viewed but not purchased. However, existing models have largely ignored the semantics of different behavior types. 
	\item \textbf{Unreasonable embedding learning}. The CMF paradigm needs to enforce the entities of one side (either users or items) have different embedding matrices for different types of behaviors. From the perspective of representation learning and interpretation of latent factor models~\cite{Zhang:2014:EFM, heckel2017scalable}, this setting is unreasonable. Specifically, a user's embedding vector represents his/her inherent interests and multiple behaviors with one item always happen in a short period. Therefore a user's embedding vector should remain unchanged when the user performs different types of behaviors on one item; and similarly for the item side. Only the interaction function~\cite{he2017neural} should be changed when predicting a user's different types of behaviors on an item.
	\item \textbf{Incapability in modeling complicated interactions}. Existing methods largely rely on MF to estimate a user's preference on an item. In MF, the interaction function is a fixed inner product, which is insufficient to model the complicated and multi-type interactions between users and items. This is also a major reason why these CMF methods need to enforce entities of one side to have different embedding matrices for predicting different types of behaviors; otherwise, the model could not make distinct predictions for different behavior types. 
	\end{itemize}

To address the above mentioned limitations in multi-behavior recommendation, we propose a new solution named \textbf{N}eural \textbf{M}ulti-\textbf{T}ask \textbf{R}ecommendation (NMTR). Briefly, our method combines the recent advance of neural collaborative filtering with multi-task learning to effectively learn from multiple types of user behaviors. Specifically, we separate the two components of embedding learning and interaction as advocated by the neural collbaborative filtering (NCF)~\cite{he2017neural} framework.
We then design that 1) a user (and an item) has a shared embedding across multiple types of behaviors, and 2) a data-dependent interaction function is learned for each behavior type. Through this way, we address the inherent limitations of CMF methods and make the model more suitable for learning from behaviors of multiple types. 

Moreover, to incorporate the behavior semantics, especially the ordinal relation among behavior types, we relate the model prediction of each behavior type in a cascaded manner. To be specific, assuming we have two form of behaviors, view and purchase, which form a natural ordinal relation: view $\rightarrow$ purchase. 
We enforce that the prediction of a high-level behavior (i.e., purchase) comes from the prediction of the low-level behavior (i.e., view). 
Through this way, we can capture the underlying semantics that a user must view a product in order to purchase it.

To summarize, the main contributions of this work are as follows. 

\begin{itemize}[leftmargin=0.5cm]
\item We propose a novel neural network model tailored to learning user preference from multi-behavior data. The model shares the embedding layer for different behavior types, and learns separate interaction function for each behavior type. 
\item To capture the ordinal relations among behavior types, we propose to correlate the model prediction of each behavior type in a cascaded way. Furthermore, we train the whole model in a multi-task manner to make full use of multiple types of behaviors. 
\item To demonstrate the effectiveness of our proposal, we implement three variants of NMTR using different neural collabrative filtering models as the interaction function. Extensive experiments on two real-world datasets show that our method outperform best existing methods by 6.08$\%$ and 30.76$\%$ on the hit-ratio effect for two datasets, respectively.
Further studies demonstrate the effectiveness of the multi-task learning manner.
\end{itemize} 

The remainder of the paper is as follows. We first formalize the problem and introduce some preliminaries in Section~\ref{sec:pre}. We then present our proposed method in Section~\ref{sec:method}. We conduct experiments in Section~\ref{sec:experiments}, before reviewing related work in Section~\ref{sec:related} and concluding the paper in Section~\ref{sec:conclusion}.  

%% file: 2.motivation.tex
\section{PRELIMINARIES}\label{sec:pre}
We first formulate the problem to solve in this paper. Then we recapitulate the neural collaborative filtering technique~\cite{he2017neural}.
Lastly, we introduce collective matrix factorization, a prevalent solution for multi-behavior recommendation.

\subsection{Problem Formulation}\label{Sec:Problem}

In recommender systems, there typically exists a key type of user behaviors to be optimized, which we term it as the \textit{target behavior}. For example, in an E-commerce site, the target behavior is usually purchase, since it is directly related with the conversion rate of recommendation and is the strongest signal to reflect a user's preference. Traditional collaborative filtering techniques~\cite{rendle2009bpr,kabbur2013fism} focus on the target behavior only and forgo other types of user behaviors such as views, clicks, etc., which are readily available in the server logs. The focus of this work is to leverage these other types of user behaviors to improve the recommendation for the target behavior. 

Let $\{\textbf{Y}^1, \textbf{Y}^2, ..., \textbf{Y}^R\}$ denote the user-item interaction matrices for all the $R$ types of behaviors. Each interaction matrix is of size $M\times N$, where $M$ and $N$ denote the number of users and items, respectively. Since in real-world applications, most user feedback are in the implicit form~\cite{rendle2009bpr,wang2017item
}, we assume that each entry of a interaction matrix has a value of 1 or 0:
\begin{equation}\label{eq:y_uir}
y^r_{ui} =\left\{
\begin{aligned}
1 & , \quad\textrm{if $u$ has interacted with $i$ under behavior $r$;} \\
0 & , \quad\textrm{otherwise.}
\end{aligned}
\right.
\end{equation}
\noindent As we have discussed in the introduction, many user behavior types in real-world applications follow an ordinal (or sequential) relationship. Without loss of generality, we assume that the behavior types have a total order and sort them from the lowest level to the highest level: $\textbf{Y}^1\rightarrow \textbf{Y}^2 ...\rightarrow \textbf{Y}^R$, where $\textbf{Y}^R$ denotes the target behavior to be optimized. 
Since the target behavior typically concerns the conversion rate, we regard it as having highest priority. 

The problem of multi-behavior recommendation is then formulated as follows.

\textbf{Input:} The user-item interaction data of the target behavior $\textbf{Y}^R$, and the interaction data of other behavior types $\{\textbf{Y}^1, \textbf{Y}^2, ..., \textbf{Y}^{R-1}\}$.

\textbf{Output:} A model that estimates the likelihood that a user $u$ will interact with an item $i$ under the target behavior. 

After obtaining the predictive model, we can use it to score all items for a user $u$, and select the top-ranked items as the recommendation results for $u$.

\subsection{Neural Collaborative Filtering (NCF)}\label{Sec:NCF}
NCF is generic neural network framework for performing collaborative filtering (CF) on single-behavior data~\cite{he2017neural}. It applies a representation learning view~\cite{RL_review} for CF, representing each user (and item) as an embedding vector. To predict a user's preference on an item, it feeds their embeddings into a neural network:
\begin{equation}
\hat{y}_{ui} = f_{\Theta} (\textbf{p}_u, \textbf{q}_i | \Theta),
\end{equation}
where $\textbf{p}_u$ and $\textbf{q}_i$ denote the embedding vector for user $u$ and item $i$, respectively; $f_{\Theta}$ denotes the neural network with parameters $\Theta$, which is also called as the \textit{interaction function}, since it is responsible for learning the interaction between user embedding and item embedding to obtain the prediction score. The model parameters of NCF can be learned in an end-to-end fashion. Specifically, the authors opt to optimize a pointwise log loss, where the positive instances are the entries of value 1 (aka., observed entries) in the user-item interaction matrix $\textbf{Y}^R$ and the negative instances are randomly sampled from the entries of value 0 (aka., missing data). 

The matrix factorization model can be seen as the special case of NCF --- by specifying the interaction function $f_{\Theta}$ as an inner product, NCF exactly recovers MF. As such, under the NCF framework, MF can be interpreted as using a fixed, data-independent interaction function. As demonstrated in the NCF paper and its follow-up work~\cite{NNCF}, using such a fixed interaction function is suboptimal and can be improved by learning the interaction function from data. It is this evidence that motivates us to develop neural network models to address the multi-behavior recommendation task. 

In the NCF paper, the authors present three instantiations of NCF, namely, GMF, MLP and NeuMF. Briefly, GMF generalizes MF by defining $f_{\Theta}$ as an element-wise product layer followed by a weighted output layer. MLP employes multi-layer perceptron above the concatenation of $\textbf{p}_u$ and $\textbf{q}_i$ to learn the interaction function. The best performance is achieved by NeuMF, which concatenates the element-wise product layer of GMF and the last hidden layer of MLP, feeding it to a weighted output layer to obtain the prediction score. Our NMTR uses NCF as a building block, and as such, any design of $f_{\Theta}$ can be used as a component to learn the interaction function for one behavior type in our method. 

\subsection{Collective Matrix Factorization}\label{Sec:CMF}
CMF is originally proposed to factorize multiple data matrices that have certain common entities~\cite{CMF}. For example, it can be used to factorize user-movie and movie-genre matrix, where movies are the common entities of the two data matrices. The idea is to correlate the multiple factorization processes by sharing the embeddings of common entities. 

Nevertheless, in multi-behavior recommendation, both sides of entities are shared in data matrices of different behavior types. Directly applying CMF will fail to produce different predictions for different behavior types. To address this problem, Zhao \textit{et al.}~\cite{zhao2015improving} proposed to share the item embedding matrix for all behavior types, allowing a user to learn different embedding vectors for different behavior types. To be specific, the objective function to optimize is as follows:
\begin{equation}
\min_{\textbf{p}_*,\textbf{q}_*} \sum_{r=1}^R \sum_{u=1}^M \sum_{i=1}^N c_{ui}^r (y_{ui}^r - {\textbf{p}_{u}^r}^T \textbf{q}_i)^2,
\end{equation}
where $c_{ui}^r$ denotes the importance of the entry $y_{ui}^r$ in factorization, $\textbf{q}_i$ denotes the embedding vector for item $i$ that is shared by all behavior types, and $\textbf{p}_u^r$ denotes the embedding vector for user $i$ in reconstructing the behaviors of the $r$-th type. Note that we have omitted the $L_2$ regularization term for clarity. 

As argued earlier in the introduction, this setting is irrational and non-interpretable as a latent factor model. Specifically, an embedding vector for a user encodes his/her latent interest, which should remain unchanged when the user seeks items of interest to consume at a particular time. 
Moreover, other potential limitations of existing CMF methods include the use of a fixed interaction function of inner product, and the use of squared regression loss for optimization, which may be suboptimal for item recommendation with implicit feedback~\cite{he2017neural,rendle2009bpr}.

%% file: 3.method.tex
\section{Methods}\label{sec:method}
Figure \ref{fig:framework} illustrates our proposed NMTR model.
Given a user-item pair $(u,i)$ as the input, the model aims to predict the likelihood that $u$ will perform a behavior (of any of the $R$ types) on item $i$, represented as the output of $\{\hat{y}_{ui}^1, \hat{y}_{ui}^2, ..., \hat{y}_{ui}^R\}$. 

Our NMTR method is featured with four special designs:

\begin{itemize}[leftmargin=0.2in]
	\item \textbf{Shared embedding layer}. To make it reasonable under the paradigm of representation learning, we share the embedding layer of users and items for the modeling of all behavior types. 
	\item \textbf{Separated interaction function}. We learn different interaction functions for predicting the behaviors of different types. This is achieved by using the expressive NCF unit for each type of behaviors. 
	\item \textbf{Cascaded predictions}. To capture the ordinal relations among behavior types, we correlate the predictions of different behavior types through cascading. 
	\item \textbf{Multi-task learning}. To optimize the cascaded architecture, we simultaneously train the predictive models for all behavior types by performing multi-task learning.
\end{itemize}

In what follows, we present our method by elaborating the above four designs.

\subsection{Shared Embedding Layer}
In order to make our proposed model extensible,
we apply one-hot encoding to encode the input of user ID and item ID. One advantage is that it can be easily extended to incorporate other features of a user and an item (e.g., user demographics and item attributes), if they are available in the application~\cite{wang2017item}. Let $\textbf{v}_u^U$ and $\textbf{v}_i^I$ denote the one-hot feature vector for user $u$ and item $i$. Then the embedding layer is defined as a linear fully connected layer without the bias terms:
\begin{equation}\label{equation:embedding}
\begin{aligned}
\textbf{p}_u = \textbf{P}^T\textbf{v}_u^U, \quad \textbf{q}_i = \textbf{Q}^T\textbf{v}_i^I, \\
\end{aligned}
\end{equation}
where $\textbf{P}$ and $\textbf{Q}$ are the user embedding matrix and item embedding matrix, respectively. When only the ID feature is used to describe a user (or an item), $\textbf{P}$ and $\textbf{Q}$ are of the size $M\times E$ and $N\times E$, respectively, where $E$ denotes the embedding size; and $\textbf{p}_u$ and $\textbf{q}_i$ are essentially the $u-$th and $i-$th row vector of $\textbf{P}$ and $\textbf{Q}$, respectively. 

It is worth noting that NMTR has only one embedding layer in the lower part of the model, which is to be used for the prediction of all behavior types in the upper part.
Although there are some works modeling dynamic interests~\cite{xiang2010temporal,rendle2010factorizing,hidasi2015session} with dynamic user embeddings, in our task it is better to use static embeddings for users. In fact, different with these works studying user dynamic intention in a long period, we focuses on modeling users multiple types of interactions on one item, which always happen in a relatively shorter time period.
 Based on this design, we can interpret the model under the paradigm of representation learning, where $\textbf{p}_u$ and $\textbf{q}_i$ are the latent features to be learned to represent user $u$ and item $i$, respectively.

\subsection{Separated Interaction Function}
Above the embedding layer is the hidden layers that model the interaction between $\textbf{p}_u$ and $\textbf{q}_i$ to obtain the prediction score. 
Since we need to predict the likelihood of multiple behavior types with the same input, it is essential to learn a separated interaction function for each type. Let $f_{\Theta}^r$ denote the interaction function for the $r$-th type of behaviors with parameters $\Theta$, which outputs the likelihood that $u$ will perform a behavior of the $r$-th type:
\begin{equation}
    \hat{y}_{ui}^r = \sigma(f_{\Theta}^r (\textbf{p}_u, \textbf{q}_i)),
\end{equation}
where $\sigma$ denotes the sigmoid function converting the output to a probability. A good design of $f_{\Theta}^r$ is to have the ability and sufficient flexibility to learn the possible complicated patterns (e.g., collaborative filtering and others) in user behaviors. 
To achieve this, we consider the three neural network units proposed in the NCF paper~\cite{he2017neural}: 
\begin{itemize}[leftmargin=*]
	\item \textbf{GMF} generalizes MF by allowing different dimensions of the embedding space to have different weights. To be specific, it first uses an element-wise product to get an interacted vector, and then project the vector to an output score with a weight vector:
	\begin{equation}
	f_{\Theta}^{GMF}(\textbf{p}_u, \textbf{q}_i) = \textbf{h}^{T}(\textbf{p}_u\odot{\textbf{q}_i}),
	\end{equation}
	where $\textbf{h}\in \mathbb{R}^{E\times 1}$ denotes the learnable weight vector. The parameters of the GMF unit are $\Theta_{GMF} = \{\textbf{h}\}$. 
	\item \textbf{MLP} applies a multi-layer perceptron on the concatenation of $\textbf{p}_u$ and $\textbf{q}_i$ to learn the interaction function in a hierarchical and non-linear manner:
	\begin{equation}\label{eq:mlp}
	\begin{aligned}
	&\textbf{z}_1 = ReLU(\textbf{W}_1 \begin{bmatrix}
	\textbf{p}_u     \\
	\textbf{q}_i       \\
	\end{bmatrix} + \textbf{b}_1), \\
	&\quad\quad\dots\dots \\
	&\textbf{z}_L = ReLU(\textbf{W}_L \textbf{z}_{L-1} + \textbf{b}_L), \\
	&f_{\Theta}^{MLP}(\textbf{p}_u, \textbf{q}_i) = \textbf{h}^T \textbf{z}_L,
	\end{aligned}
	\end{equation}
	where $L$ denotes the number of hidden layers in the multi-layer perceptron, $\textbf{W}_x$ and $\textbf{b}_x$ denote the weight matrix and bias vector for the $x$-th hidden layer, and $\textbf{z}_x$ are the intermediate neurons. By default, the rectifier unit (ReLU) is used as the activation function for the hidden layer, which is beneficial to build deep models. The parameters of the MLP unit are $\Theta_{MLP} = \{\textbf{h}, \{\textbf{W}_x\}_{x=1}^L, \{\textbf{b}_x\}_{b=1}^L \}$.
	\item \textbf{NeuMF} combines the advantage of the linear GMF with the nonlinear MLP to learn the interaction function:
	\begin{equation}
	\begin{aligned}
	f_{\Theta}^{NeuMF}(\textbf{p}_u, \textbf{q}_i) = \textbf{h}^T\begin{bmatrix}
	\textbf{p}_u\odot{\textbf{q}_i}      \\
	\textbf{z}_L      \\
	\end{bmatrix},
	\end{aligned}
	\end{equation}
	where $\textbf{z}_L$ indicates the last hidden layer of MLP, as have been defined in Equation~(\ref{eq:mlp}). 
	The $\textbf{z}_L$ is concatenated with $\textbf{p}_u\odot{\textbf{q}_i}$ as the hidden layer of NeuMF, which is then projected to a score through the weighted vector $\textbf{h}\in\mathbb{R}^{2E\times 1}$.
	In the original design of NeuMF, the authors used different embedding layers for GMF and MLP. While in our method, we have only one set of embeddings for users and items. As such, we tweak the NeuMF unit by sharing the embedding layer of GMF and MLP.
\end{itemize}
Note that any of the three units can be used to model for behaviors of any type, and the optimal setting may depend on the dataset. We will empirically evaluate the performance of three NCF choices and their impact on our NMTR method in Section~\ref{sec:experiments}.
There are many other possible designs for the NCF units, such as placing more layers above the hidden layer of NeuMF to thoroughly merge GML and MLP, among others~\cite{NNCF,wang2017item}. Since the focus of this paper is not to develop new NCF units for interaction learning, we leverage existing ones as the building block for our NMTR model. 

\begin{figure}[t]\label{fig:framework}
	\includegraphics[width=0.50\textwidth]{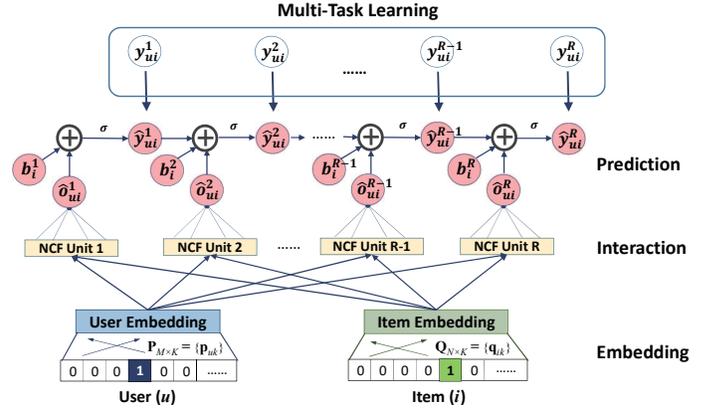}	
	\caption{Illustration of our proposed NMTR model.}
	\label{fig:framework}
\end{figure}

\subsection{Cascaded Predictions}
Typically there are certain ordinal relations among behavior types in a real-world application, such as a user must view a product (i.e., click the product page) before she can purchase it. 
The existence of such relations implies that the predictive models for different behavior types should be related with each other, rather than being independent. 
To encode the sequential effect, we enforce that the prediction on a behavior type lies in the predictions of the precedent behavior types. Formally, we cascade the predictions of different behaviors as:
 \begin{equation}\label{equation:cascade}
 \begin{aligned}
  \vspace{0.1cm}
 \hat{y}^R_{ui} &= \sigma(\hat{y}^{R-1}_{ui} + f_\Theta^R(\textbf{p}_u, \textbf{q}_i) + b^R_i), \\
 &\quad\quad\cdots\cdots \\
 \hat{y}^{2}_{ui} &= \sigma(\hat{y}^{1}_{ui} + f_\Theta^{2}(\textbf{p}_u, \textbf{q}_i) + b^{2}_i), \\
 \hat{y}^{1}_{ui} &= \sigma(f_\Theta^{1}(\textbf{p}_u, \textbf{q}_i) + b^{1}_i), \\
 \vspace{0.0cm}
 \end{aligned}
 \end{equation}
where $b_i^r$ denotes the bias of item $i$ in the data of the $r$-th behavior type, and $f_\Theta^r$ denotes the interaction function for the $r$-th type of behaviors, which can be any of the three NCF units as introduced before.
The item bias term can capture some discrepancy effects in different types of behaviors, for example, some items are likely to be clicked by users (e.g., products on campaign) but less likely to be purchased. 
Moreover, some previous work has demonstrated that incorporating item bias is more effective than incorporating user bias for learning from single-behavior implicit feedback~\cite{kabbur2013fism}. 

A graphical illustration of our cascading design\footnote{Note that we assume that the behaviors can form a full-order cascading relationship, while in real world the relationship might be more complicated. For example, there is no sequential relation between sharing a product to social network and adding it to cart by nature. Technically speaking, we can adapt to such partial-order relation by sorting the behaviors by their strength in reflecting user preference. We leave this exploration as future work. } can be found in the top part of Figure~\ref{fig:framework}. Such a design is particularly useful for predicting the preference of inactive users that have few data on the target behavior. Typically, the data of low-level behaviors (e.g., clicks) is easier to collect and has a larger volume than the target behavior (e.g., purchases). By basing the prediction of target behavior on its precedent types of behaviors, we can achieve better prediction when the target behavior data of a user is insufficient to estimate $f_\Theta^R$ well.

\subsection{Multi-Task Learning}\label{Sec:MTL}
As we have a dedicated model for each type of behaviors and the models follow a cascading prediction, it is intuitive to train models separately by following the order of $\hat{y}_{ui}^1, \hat{y}_{ui}^2, ..., \hat{y}_{ui}^R$. 
Since these models share the same embedding layer and the final recommendation is based on the last target model $\hat{y}_{ui}^R$, this way can be seen as pre-training the embedding layer of the target model using other types of behaviors. We argue that such a sequential training manner does not make full use of the multi-behavior data, since it only uses precedent models to improve the next model while there is no benefit for the precedent models. A better solution could be to let the models reinforce each other.

In contrast to training the models separately, multi-task learning (MTL) is a paradigm that performs joint training on the models of different but correlated tasks, so as to obtain a better model for each task~\cite{evgeniou2004regularized}. The intuition for our design of cascaded predictions is that, if we can obtain improved models for other types of behavior, the model for the target behavior can also be improved. As such, we opt for MTL that trains all models simultaneously, where the model learning for each behavior type is treated as a task. \vspace{+5pt}

\noindent \textbf{Objective Function}. Following the probabilistic optimization framework~\cite{he2017neural}, we first define the likelihood function for a single behavior type as:
\begin{equation}
P_r = \prod_{(u,i)\in{\mathcal{Y}_r^+}}{\hat{y}_{ui}^r} \prod_{(u,i)\in{\mathcal{Y}_r^-}}{{(1-\hat{y}_{ui}^r)}}, 
\end{equation}
where $\mathcal{Y}_r^+$ denotes the set of observed interactions in behavior matrix $\textbf{Y}^r$, and  $\mathcal{Y}_r^-$ denotes negative instances to be sampled from the unobserved interactions in  $\textbf{Y}^r$. 
We then get the joint probability for multiple types of behaviors as:
\begin{equation}
    P = \prod_{r=1}^R P_r = \prod_{r=1}^R \prod_{(u,i)\in{\mathcal{Y}_r^+}}{\hat{y}_{ui}^r} \prod_{(u,i)\in{\mathcal{Y}_r^-}}{{(1-\hat{y}_{ui}^r)}}.
\end{equation}

Taking the negative logarithm of the joint probability, we obtain the loss function to be minimized as:
\begin{equation}
    L = -\sum_{r=1}^{R}{\lambda_r(\sum_{(u,i)\in{\mathcal{Y}{^+_r}}}{\log{\hat{y}_{ui}^r}}  + \sum_{(u,i)\in{\mathcal{Y}{^-_r}}}{\log(1 - {\hat{y}_{ui}^r})}) },
\end{equation}
where we additionally include the term $\lambda_r$ to control the influence of the $r$-th type of behaviors on the joint training. This is a hyper-parameter to be specified for different datasets, since the importance of a behavior type may vary for problems of different domains and scales. We additionally enforce that $\sum_{r=1}^R \lambda_r = 1$ to facilitate the tuning of these hyper-parameters. 

Directly optimizing this joint loss function will update the parameters of models for multiple behavior types together. As such, a better embedding learned from a gradient step of the data of one type will benefit the learning of other types.  \vspace{+5pt}

\noindent \textbf{Training}. Since our model is composed of nonlinear neural networks, we optimize parameters with stochastic gradient descent (SGD), a generic solver for neural network models. As most machine learning toolkits (e.g., TensorFlow, Theano, PyTorch etc.) provide the function of automatic differentiation, we omit the derivation of the derivatives of our model. Instead, we elaborate on how to form a mini-batch to facilitate faster training, since modern computing units like GPU and CPU provide acceleration for matrix-wise float operations. 

To generate a mini-batch, we first sample a user-item pair $(u,i)$ such that user $u$ has at least one observed interaction on item $i$ (regardless of the behavior type). We then inspect the interactions of the $(u,i)$ pair --- for each observed interaction, we sample a negative instance from $u's$ unobserved interactions of the behavior type. As an example, if the sampled $(u,i)$ pair has an interaction in the 1-st behavior and 2-nd behavior, we get two positive training instances $y_{ui}^1$ and $y_{ui}^2$; we then sample two items $t$ and $s$ that $u$ did not interact under first two behaviors, respectively, to get two negative instances $y_{ut}^1$ and $y_{us}^2$. We iterate the above sampling step until the desired size of a mini-batch is reached. 

Note that we empirically find that sampling multiple negative instances to pair with a positive instance in a mini-batch can improve the performance. This finding has been reported before in optimizing neural recommender models with log loss on single-behavior data~\cite{he2017neural,DBLP:conf/sigir/ChenZAXYQ17}. As such, in our experiments, we allow a flexible tuning of the negative sampling ratio.

%% file: 4.evaluation.tex
\section{Experiments}\label{sec:experiments}

In this section, we conduct extensive experiments on two real-world datasets to answer the following research questions:

\begin{itemize}[leftmargin=*]
	
	\item \textbf{RQ1:} How does our proposed NMTR perform as compared with state-of-the-art recommender systems that are designed for learning from single-behavior and multi-behavior data?
	
	\item \textbf{RQ2:} How do the key hyper-parameters affect NMTR's performance, and how is the effectiveness of our designed multi-task learning for the task? 
	\item \textbf{RQ3:} Can NMTR help to address the data sparsity problem, i.e., improving recommendations for sparse users with fewer interactions of the target behavior?
\end{itemize}

\noindent In what follows, we first describe the experimental settings, and then answer the above three research questions.

\subsection{Experimental Settings}\label{Sec:ExpSet}

\subsubsection{Datasets and Evaluation Protocol}\label{section:dataset}
We experimented with two real-world E-commerce datasets that contain multiple types of user behaviors including purchases, views, adding to carts, etc.
\begin{itemize}[leftmargin=*]
	\item \textbf{Beibei Dataset\footnote{https://www.beibei.com}.} 
	This dataset is collected from Beibei, the largest E-commerce platform for maternal and infant products in China. We sampled a subset of user interactions that contain views, adding to carts (abbreviated as \textit{carts}), and purchases within the time period from 2017/06/01 to 2017/06/30. 
	\item \textbf{Tmall Dataset\footnote{https://www.tmall.com}.}
	This is the dataset released in IJCAI-15 challenge\footnote{https://tianchi.aliyun.com/datalab/dataSet.htm?id=5}, which is collected from Tmall, the largest business-to-consumer E-Commerce website in China. It records two types of user behaviors, views and purchases, within the time period from 2014/05/01 to 2014/11/30. 
\end{itemize}

For both datasets, we merged the duplicated user-item interactions by keeping the earliest one; the rationality here is to test the performance of a method in recommending novel items that a user did not consume before. 
Moreover, we focused on users with more than one type of behavior. 
After the above pre-processing steps, we obtained the final evaluation datasets, the statistics of which are summarized in Table~\ref{tab:Dataset}. For these two datasets, there exist strict cascading relationships. For example, in Beibei dataset, a user must click first before adding to cart, and must add to cart first before purchasing.
In the evaluation stage, given a user in the testing set, each algorithm ranks all items that the user has not interacted before. 
We applied the widely used leave-one-out technique to obtain the training set and test set, which means for every user, there is a test item her has not interacted with.
We then adopted two popular metrics, \emph{HR} and \emph{NDCG}, to judge the performance of the ranking list:
\begin{itemize}[leftmargin=*]
	\item \textbf{HR@K:}
	\emph{Hit Ratio} (HR) measures whether the test item is contained by the top-K item ranking list (1 for yes and 0 for no).
	\item \textbf{NDCG@K:} 
	\emph{Normalized Discounted Cumulative Gain} (NDCG) complements HR by assigning higher scores to the hits at higher positions of the ranking list.
\end{itemize}

\begin{table}[t]
	\begin{center}
		\begin{adjustwidth}{0.4cm}{}
			\caption{Statistics of our evaluation datasets.}\label{tab:Dataset}
			\begin{tabular}{|c|c|c|c|c|c|}
				\hline
				{\bf Dataset}    &  {\bf User\#}     &  {\bf Item\#}    &{\bf Purchase\#}  &{\bf Cart\#}  &{\bf View\#}   \\\hline
				Beibei       & 21,716  & 7,977 & 295,622 & 642,622 & 2,412,586   \\\hline
				Tmall            & 15,670  & 9,076  & 136,648 &  -- & 813,396  \\\hline
			\end{tabular}
		\end{adjustwidth}
	\end{center}
\end{table}

\begin{table}[t]
\centering
\scriptsize
\caption{Best parameter settings of our proposed NMTR methods for top-K recommendation}
\label{tab:Parameter}
\begin{adjustwidth}{-0.4cm}{}
\begin{tabular}{|c|c|c|c|c|}
\hline
\textbf{Dataset}                                                                   & \textbf{Parameter}           & \textbf{NMTR-GMF} & \textbf{NMTR-MLP} & \textbf{NMTR-NeuMF} \\ \hline
\multirow{5}{*}{\textbf{\begin{tabular}[c]{@{}c@{}}Beibei\end{tabular}}} & \textbf{Optimzer}            & Adagrad           & Adagrad           & Adagrad             \\ \cline{2-5} 
                                                                                   & \textbf{Learning rate}       & 0.01              & 0.01              & 0.01                \\ \cline{2-5} 
                                                                                   & \textbf{Number of layer}     & -                 & 3                 & 3                   \\ \cline{2-5} 
                                                                                   & \textbf{Loss coefficient}    & {[}1/3,1/3,1/3{]} & {[}1/3,1/3,1/3{]} & {[}1/3,1/3,1/3{]}   \\ \cline{2-5} 
                                                                                   & \textbf{Regularization} & {[}0,1e-5{]}      & {[}0,1e-5{]}      & {[}0,1e-5{]}        \\ \hline
\multirow{5}{*}{\textbf{\begin{tabular}[c]{@{}c@{}}Tmall\end{tabular}}}  & \textbf{Optimzer}            & Adagrad           & Adagrad           & Adagrad             \\ \cline{2-5} 
                                                                                   & \textbf{Learning rate}       & 0.01              & 0.01              & 0.05                \\ \cline{2-5} 
                                                                                   & \textbf{Number of layer}     & -                 & 3                 & 3                   \\ \cline{2-5} 
                                                                                   & \textbf{Loss coefficient}    & {[}0.4,0.6{]}     & {[}0.5,0.5{]}     & {[}0.4,0.6{]}       \\ \cline{2-5} 
                                                                                   & \textbf{Regularization term} & {[}0,5e-5{]}      & {[}0,1e-5{]}      & {[}0,0{]}           \\ \hline
\end{tabular}
\end{adjustwidth}
\end{table}

\subsubsection{Baselines}

We compared the performance of our proposed NMTR with 9 baselines, which can be divided into two groups based on whether it models single-behavior or multi-behavior data. 
The compared single-behavior methods are introduced as follows.

\para{BPR~\cite{rendle2009bpr}} \emph{Bayesian Personalized Ranking} (BPR) is a widely used pairwise learning framework for item recommendation with implicit feedback. Same as the original paper, we used BPR to optimize the MF model.
\para{NCF~\cite{he2017neural}} \emph{Neural Collaborative Filtering} (NCF) is a neural framework to learn interactions between the latent features of users and items. As we employed three NCF methods, named  \textbf{GMF}, \textbf{MLP} and \textbf{NeuMF} to learn the interaction function for each behavior type, we evaluated how the three methods perform for single-behavior data. 

\vspace{0.2cm}
\noindent The second group of five compared methods that can leverage multiple types of behavior data are as follows. 

\para{CMF~\cite{zhao2015improving}}
As have described in Section~\ref{Sec:CMF}, CMF decomposes the data matrices of multiple behavior types simultaneously. 
We adapted the method by sharing the user embeddings for factorizing different interaction matrices of various types of behaviors.
As our datasets are implicit feedback, we further augmented the method by sampling negative instances in the same way as our NMTR.

\para{MC-BPR~\cite{loni2016bayesian}}
Multi-Channel BPR \cite{loni2016bayesian} is the state-of-the-art solution for multi-behavior recommendation. It adapts the negative sampling rule in BPR to account for the levels of user feedback in multi-behavior data. 
For example on the Tmall dataset that has two behavior types --- purchase and view, 
to generate a negative sample for a purchase interaction, it assigns different probabilities for sampling from 1) items that are viewed but not purchased, and 2) items that are not viewed. We tuned the probability distribution for sampling and reported the best results.

\para{MC-NCF} Since Multi-Channel BPR is a generic learning method that is applicable to any differentiable recommender model, we replaced the basic MF model in it with state-of-the-art NCF models, and named this extension as MC-NCF. That is, we optimized the three NCF models with the Multi-Channel BPR learner, and named the respective methods as \textbf{MC-GMF}, \textbf{MC-MLP} and \textbf{MC-NeuMF}.

\begin{table*}[h]
	\begin{adjustwidth}{-0cm}{}
	\footnotesize
		\begin{center}
			\caption{Top-K recommendation performance comparison on the Beibei and Tmall datasets (K is set to 50, 80, 100, 200) }\label{tab:Comparision}
			\begin{tabular}{|c|c|c|c|c|c|c|c|c|c|c|}
				\hline			
				
				\multicolumn{2}{|c|}{}    & \multicolumn{8}{|c|}{\bf Beibei Dataset}   \\\hline
				\bf Group&\bf Method     &\bf HR@50  & \bf NDCG@50 & \bf HR@80 & \bf NDCG@80 &\bf HR@100 & \bf NDCG@100 &\bf HR@200 & \bf NDCG@200 \\\whline
				
				\multirow{3}{*}{Our NMTR Model}&NMTR-GMF     &0.2050 &0.0590 &\bf 0.2721 &\bf 0.0688 &0.3119 &0.0741 &0.4543 &0.0961 \\\cline{2-10}
				&NMTR-MLP     &0.1928 &0.0560 &0.2690 &0.0676 &0.3188 &\bf 0.0762 &0.4732 &0.0967 \\\cline{2-10}
				&NMTR-NeuMF   &\bf 0.2079 &\bf 0.0609 &0.2689 &0.0683 &\bf 0.3193  & 0.0760 &\bf 0.4766 &\bf 0.0971 \\\whline
				
				\multirow{5}{*}{Multi-behavior}&CMF     &0.1596 &0.0481 &0.2377 &0.0606 &0.2829 &0.0663 &0.4191 &0.0850  \\\cline{2-10}
				
				&MC-BPR  &0.1743 &0.0503 &0.2299 &0.0604 &0.2659 &0.0647 &0.3852 &0.0786 \\\cline{2-10}
				&MC-GMF  &0.1822 &0.0508 &0.2425 &0.0611 &0.2975 &0.0690 &0.4262 &0.0891 \\\cline{2-10}		
				&MC-MLP  &0.1810 &0.0534 &0.2342 &0.0598 &0.2810 &0.0684 &0.4116 &0.0834 \\\cline{2-10}
				&MC-NeuMF&0.2014 &0.0577 &0.2522 &0.0669 &0.3010 &0.0719 &0.4300 &0.0897 \\\whline
				
				\multirow{4}{*}{Single-behavior}&BPR   &0.1199 &0.0348 &0.1686 &0.0419 &0.2002 &0.0463 &0.3039 &0.0624\\\cline{2-10}
				&GMF   &0.1792 &0.0475 &0.2555 &0.0608 &0.2920 &0.0665 &0.4090 &0.0828\\\cline{2-10}
				&MLP   &0.1711 &0.0483 &0.2383 &0.0459 &0.2679 &0.0617 &0.3947 &0.0792\\\cline{2-10}
				&NeuMF &0.1828 &0.0573 &0.2559 &0.0668 &0.2929 &0.0714 &0.4078 &0.0852\\\cline{1-10}

				\hline       \hline

			\multicolumn{2}{|c|}{}    & \multicolumn{8}{|c|}{\bf Tmall Dataset}   \\\hline
				\bf Group&\bf Method     &\bf HR@50  & \bf NDCG@50 & \bf HR@80 & \bf NDCG@80 &\bf HR@100 & \bf NDCG@100 &\bf HR@200 & \bf NDCG@200 \\\whline
				
				\multirow{3}{*}{Our NMTR Model}
				&NMTR-GMF   &0.0778 &0.0250 &0.1042 &0.0293 &\bf0.1196 &0.0314 &\bf0.1751 &0.0390\\\cline{2-10}
				&NMTR-MLP   &0.0734 &0.0251 &0.0884 &0.0277 &0.0982 &0.0290 &0.1672 &0.0338\\\cline{2-10}
				&NMTR-NeuMF  &\bf0.0854 &\bf0.0315 &\bf0.1045 &\bf0.0347 &0.1169 &\bf0.0366 &0.1668 &\bf0.0428\\\whline
				
				\multirow{5}{*}{Multi-behavior}
				&CMF                &0.0738 &0.0234 &0.0940 &0.0269 &0.1085 &0.0287 &0.1565 &0.0359 \\\cline{2-10}
				&MC-BPR  &0.0674 &0.0218 &0.0928 &0.0260 &0.1072 &0.0282 &0.1597 &0.0357 \\\cline{2-10}
				&MC-GMF  &0.0653 &0.0243 &0.0778 &0.0258 &0.0846 &0.0264 &0.1084 &0.0294\\\cline{2-10}
				
				&MC-MLP &0.0617 &0.0195 &0.0784 &0.0219 &0.0868 &0.0228 &0.1122 &0.0238\\\cline{2-10}
				&MC-NeuMF &0.0711 &0.0296 &0.0820 &0.0311 &0.0894 &0.0320 &0.1172 &0.0359\\\whline
				
				\multirow{4}{*}{Single-behavior}&BPR      &0.6666 &0.0200 &0.0926 &0.0240 &0.1058 &0.0263 &0.1647 &0.0342\\\cline{2-10}
				&GMF     &0.0742 &0.0271 &0.0927 &0.0295 &0.1027 &0.0306 &0.1407 &0.0355 \\\cline{2-10}
				&MLP     &0.0666 &0.0194 &0.0824 &0.0220 &0.0905 &0.0233 &0.1194 &0.0273\\\cline{2-10}
				&NeuMF   &0.0760 &0.0299 &0.0925 &0.0321 &0.1013 &0.0333 &0.1383 &0.0377\\\cline{1-10}

			\end{tabular}
			
		\end{center}
	\end{adjustwidth}
	\vspace{0.3cm}
\end{table*}

\begin{figure*}[h]
	\centering
	\subfigure[Training Loss]{
		\label{fig:overall1-loss}
		\includegraphics[width=0.31\textwidth]{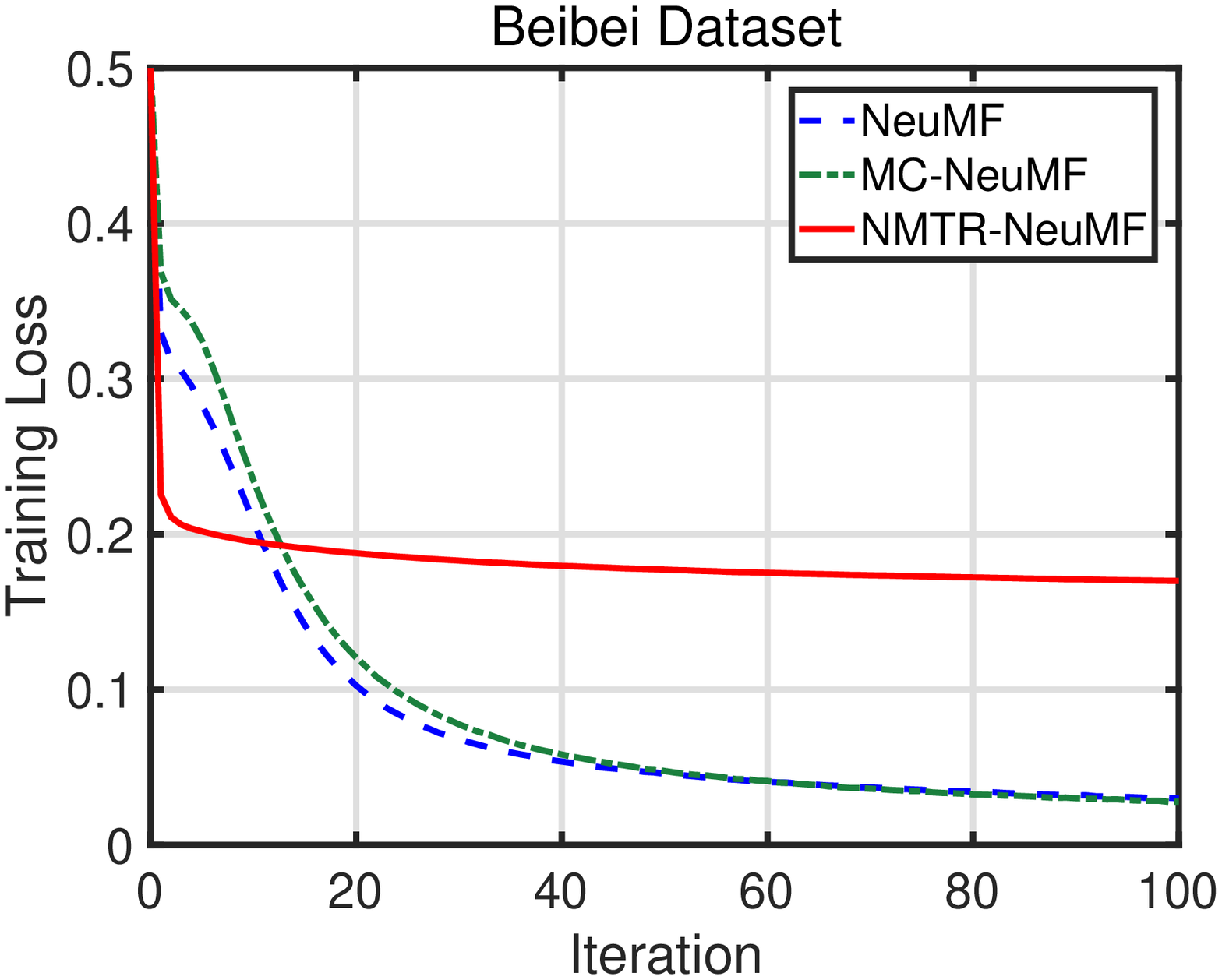}}
	\subfigure[HR@100]{
		\label{fig:overall1-hr}
		\includegraphics[width=0.31\textwidth]{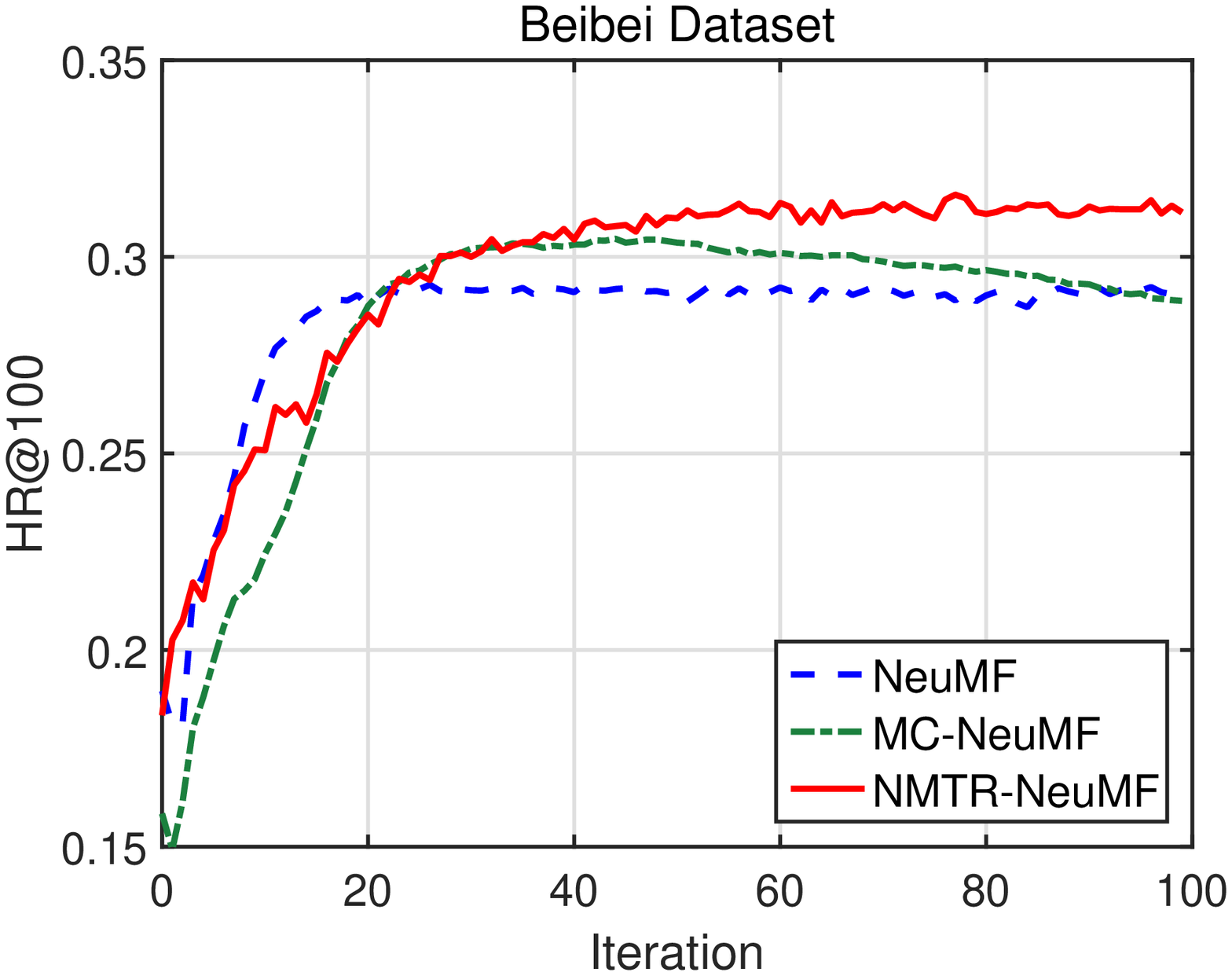}}
	\subfigure[NDCG@100]{
		\label{fig:overall1-ndcg}
		\includegraphics[width=0.31\textwidth]{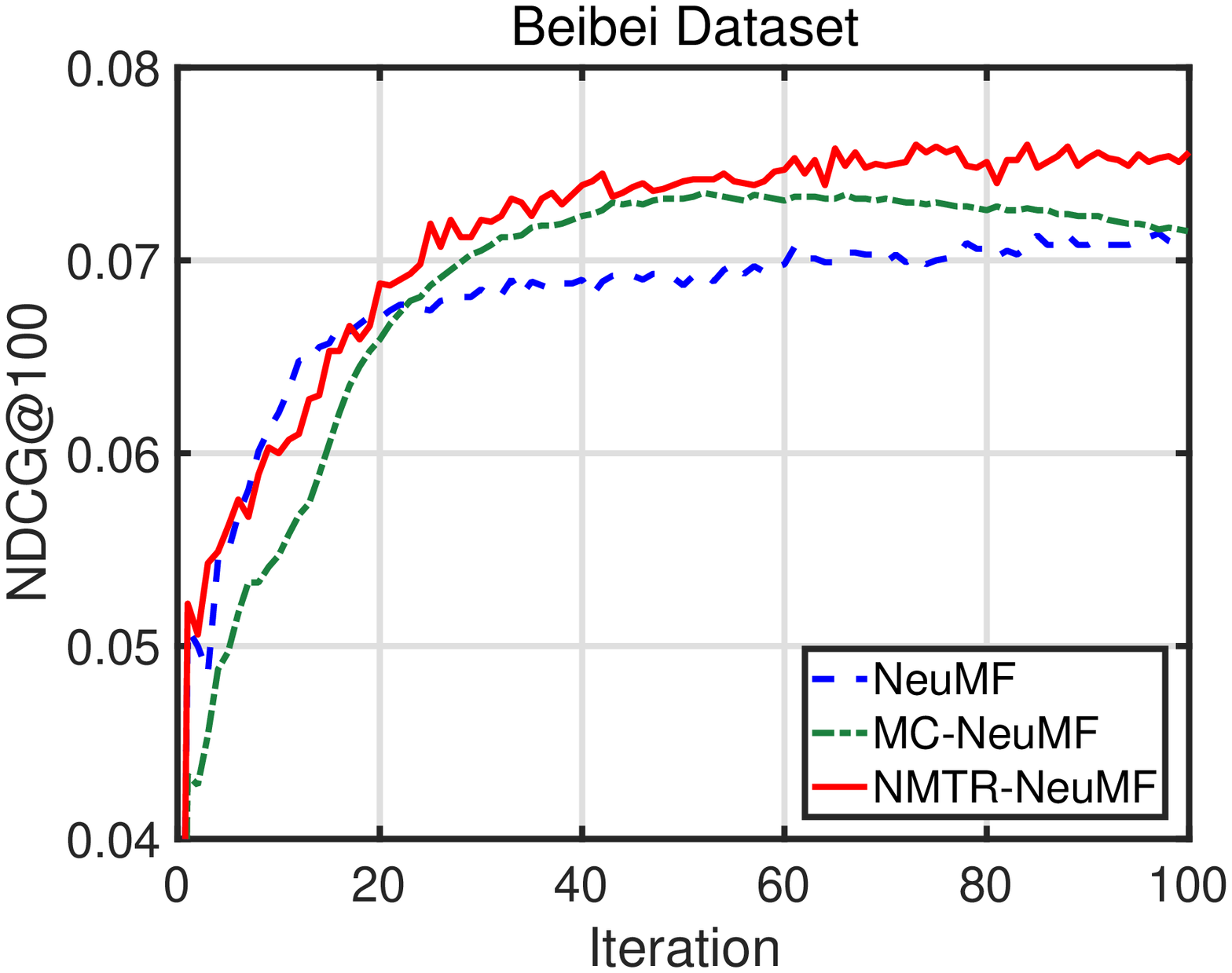}}
	\caption{Training loss and testing performance of NeuMF, MC-NeuMF, and NMTR-NeuMF in each iteration on Beibei}
	\label{fig:overall1}
	
\end{figure*}

\begin{figure*}[h]
	\centering
	\subfigure[Training Loss]{
		\label{fig:overall-top-1}
		\includegraphics[width=0.31\textwidth]{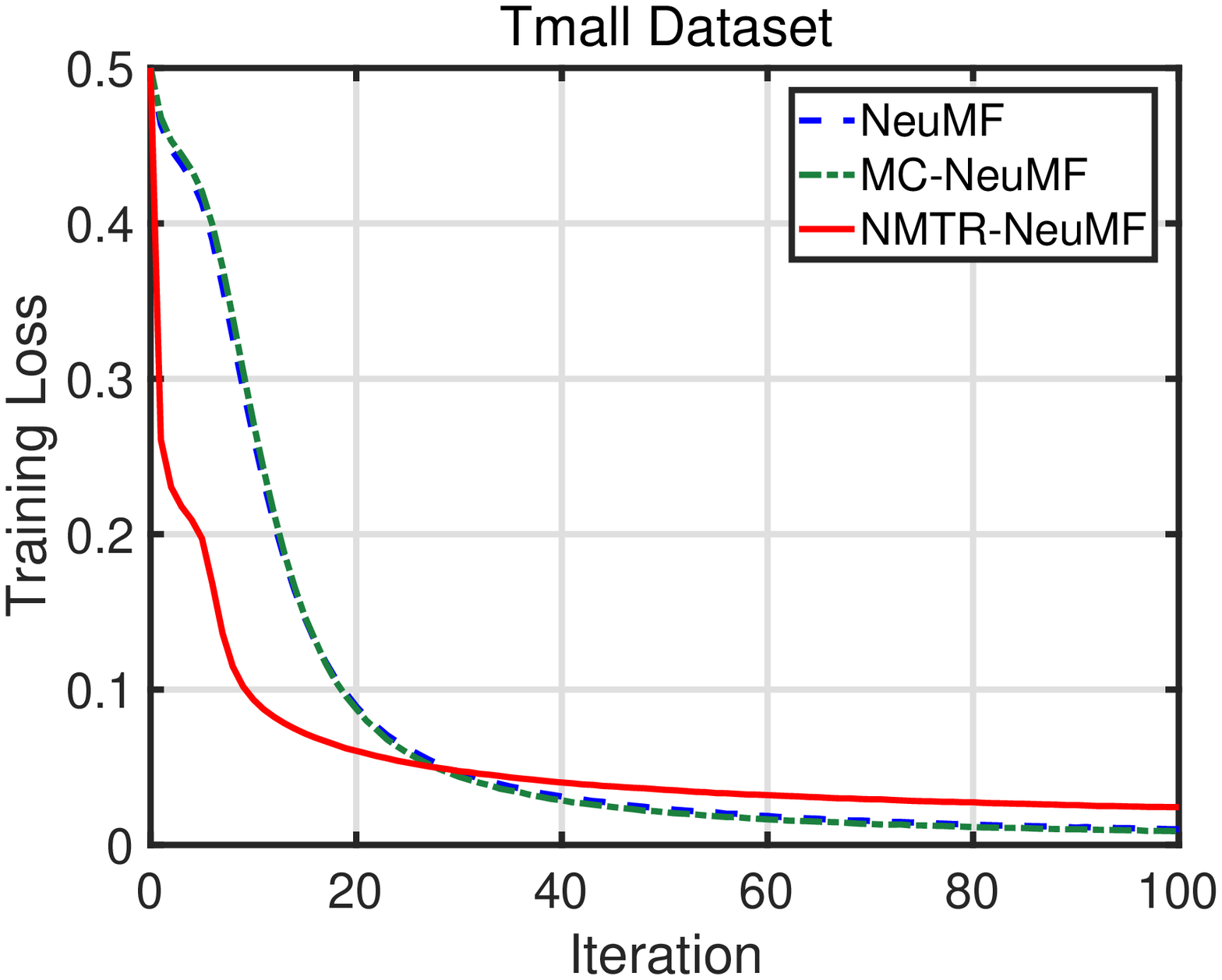}}
	\subfigure[HR@100]{
		\label{fig:overall-top-5}
		\includegraphics[width=0.31\textwidth]{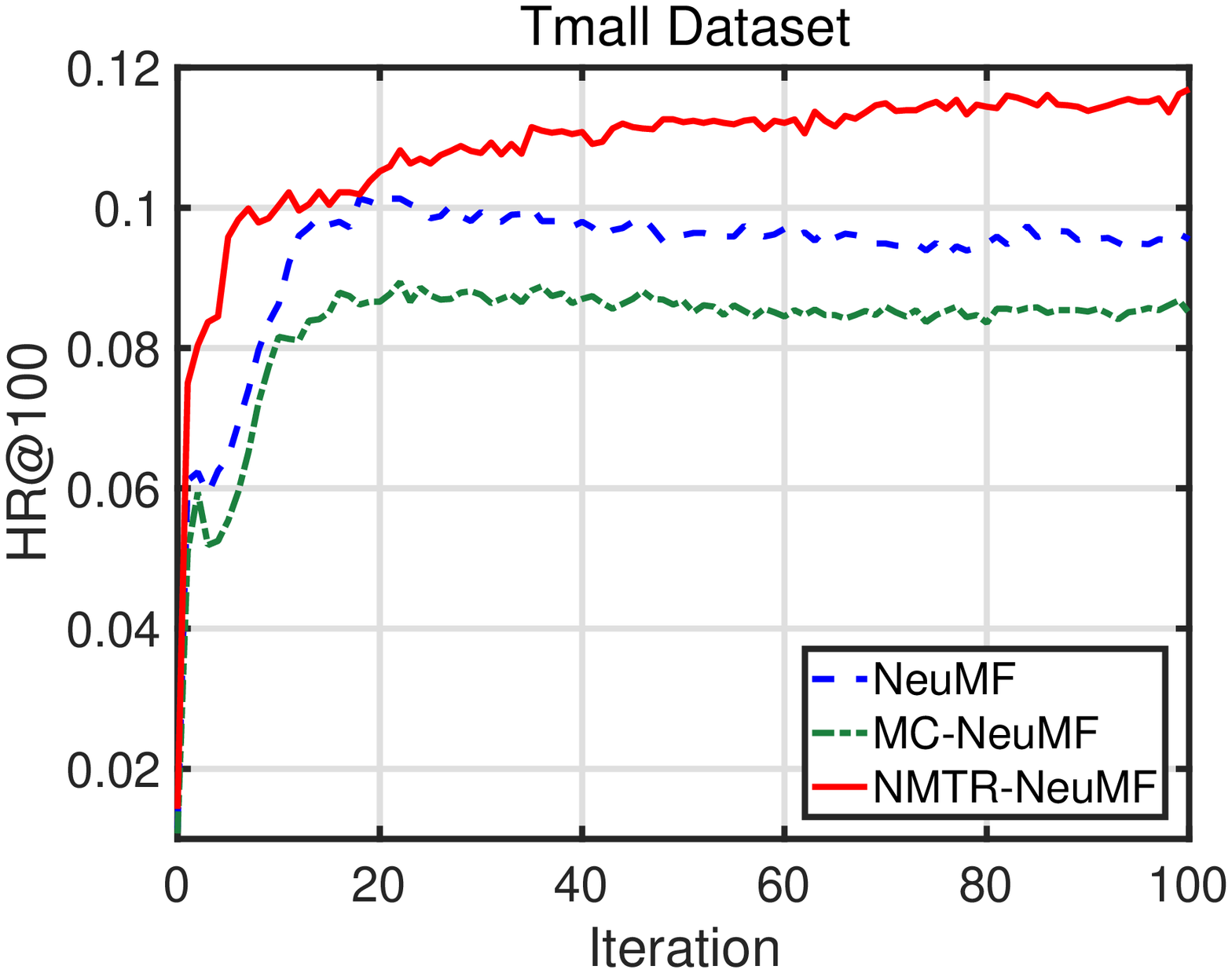}}
	\subfigure[NDCG@100]{
		\label{fig:overall-top-5}
		\includegraphics[width=0.31\textwidth]{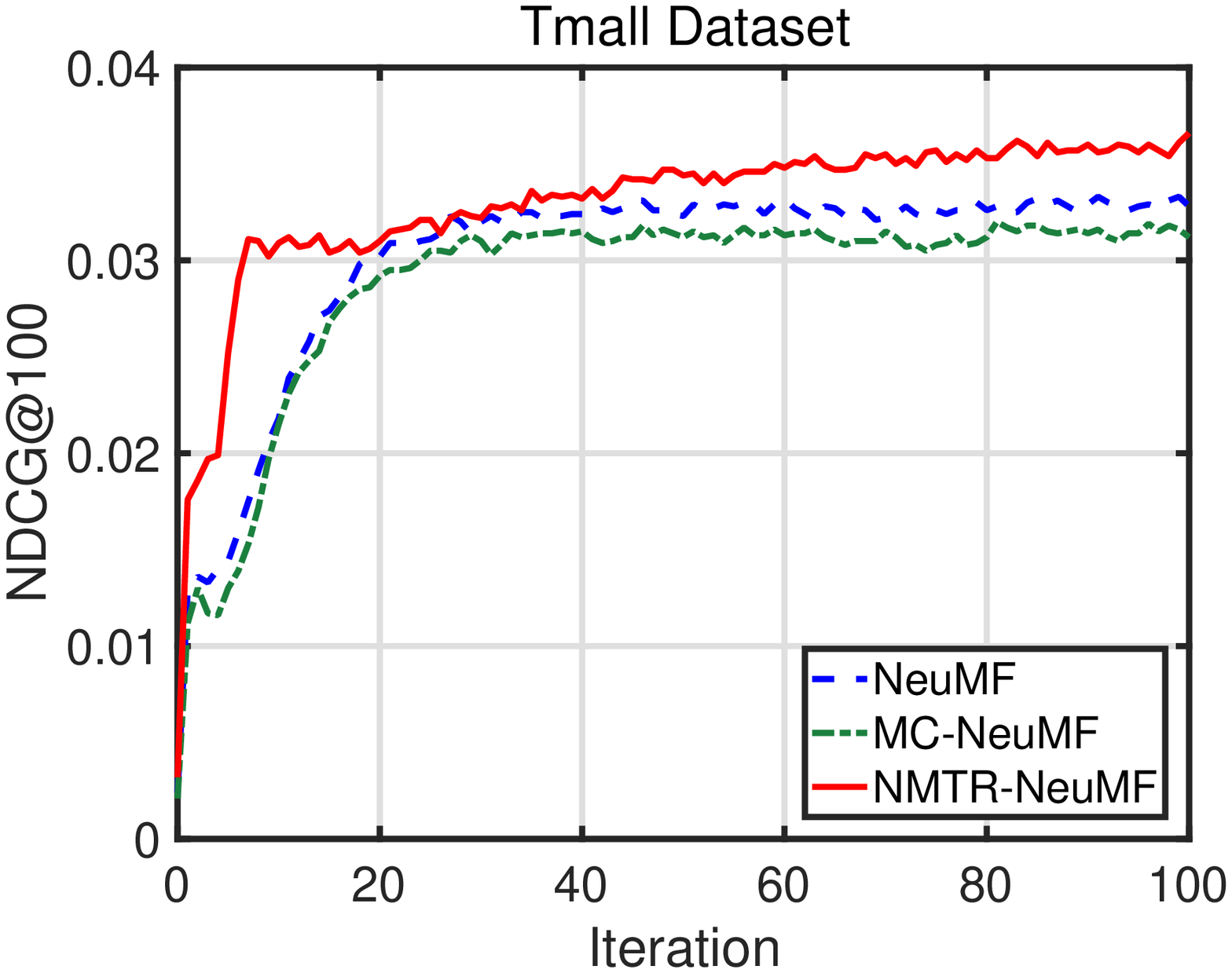}}
	\caption{Training loss and testing performance of NeuMF, MC-NeuMF, and NMTR-NeuMF in each iteration on Tmall}
	\label{fig:overall2}
\end{figure*}

\begin{table}[t]
	\begin{center}
		\begin{adjustwidth}{0.4cm}{}
			\caption{Average training time of one epoch of NeuMF, MC-NeuMF and NMTR-NeuMF on two datasets.}\label{tab:trainingtime}
			\begin{tabular}{|c|c|c|c|c|c|}
				\hline
				{\bf Dataset}    &  {\bf NMTR-NeuMF\#}     &  {\bf MC-NeuMF\#}    &{\bf NeuMF\#}   \\\hline
				Beibei       & 204.4s  &472.4s & 101.9s    \\\hline
				Tmall            & 91.5s  & 73.1s  & 88.0s  \\\hline
			\end{tabular}
		\end{adjustwidth}
	\end{center}
\end{table}

\subsubsection{Parameter Settings}

We implemented our NMTR\footnote{We release our implementation along with datasets at https://github.com/fiblab} and baseline methods in TensorFlow\footnote{https://www.tensorflow.org}. 
Since we have three choices of NCF units as the interaction function, we name the respective methods as \textbf{NMTR-GMF}, \textbf{NMTR-MLP} and \textbf{NMTR-NeuMF}.
We randomly selected a training instance for each user as the validation set to tune hyper-parameters. 
For all methods, we set the embedding size to 64, a relatively larger number that achieves good performance on our datasets. 
For CMF, one important hyper-parameter is the weight of different behavior types in the joint loss. We tuned the weight for each behavior in [0, 0.2, 0.4, 0.6, 0.8, 1]. To be specific, the weight for each behavior represent the influence of each interaction matrix on the collective matrix factorization task. For MC-methods, we carefully tune the sampling distribution following the original paper.
For neural network models, 
we initialized their parameters using the method proposed in~\cite{he2015delving}.
For models that have multiple hidden layers, i.e., MLP, MC-MLP, NMTR-MLP, NeuMF, MC-NeuMF and NMTR-NeuMF, we employed a tower structure for the hidden layers same as ~\cite{he2017neural}, and tuned the number of layers from 1 to 5. 
We set the negative sampling ratio as 4 for all methods, an empirical value that shows good performance. 
We tried two SGD-based optimizers, Adam~\cite{kingma2014adam} and Adagrad~\cite{duchi2011adaptive}, and tuned the learning rate for each optimizer in [0.001, 0.005, 0.01, 0.02, 0.05]. 
Moreover, we applied $L_2$ regularization to all methods to prevent overfitting.

\subsection{Performance Comparison (RQ1)}

We first compare the top-K recommendation performance with state-of-the-art methods. 
We investigate the top-$K$ performance with $K$ setting to  [50, 80, 100, 200]. 
Note that for a user, our evaluation protocol ranks all unobserved items in the training set~\cite{kabbur2013fism}. 
Though this all-ranking protocol can be very time-consuming, the obtained results are more persuasive than ranking a random subset of negative times only (e.g., as have done in~\cite{he2017neural}). 
In this case, small values of $K$ will make the results have a large variance and unstable. As such, we report results of a relatively large\footnote{There is another reason to choose a relatively larger $K$. In practical recommender systems, the procedure of item recommendation is typically divided into two stages~\cite{wang2016spore}: candidate selection and re-ranking. Since collaborative filtering (CF) methods are typically applied in the first stage to retrieve a few hundreds of relevant items, a larger $K$ to evaluate CF methods is more reasonable.
}.
We report the best parameter setting for our proposed NMTR methods in Table~\ref{tab:Parameter}.

Table~\ref{tab:Comparision} shows the performance of HR@K and NDCG@K for our three NMTR methods, five multi-behavior recommendation methods, and four single-behavior methods. From the results, we have the following observations:
\begin{itemize}[leftmargin=*]
	\item \textbf{NMTR achieves the best performance.} Our proposed NMTR methods obtain the best performance in terms of HR@K and NDCG@K as compared to all baselines. 
	The one-sample paired t-tests indicate that all improvements are statistically significant for $p$ $<$ 0.05.
	Among the three NMTR methods, NMTR-GMF and NMTR-NeuMF are better than NMTR-MLP, which verifies the effectiveness of the element-wise operator in learning the user-item interaction function. 
	
	Compared with the best single-behavior baseline NeuMF, NMTR outperforms it by 9.01\% in HR and 6.72\% in NDCG on the Beibei dataset; and the improvements are 13.04\% in HR and 9.91\% in NDCG on the Tmall dataset.
	Compared with MC-NeuMF, which extends NeuMF on multi-behavior data with the Multi-Channel BPR~\cite{loni2016bayesian}, NMTR obtains an improvement in HR of 6.08\% and 10.23\% on Beibei and Tmall, respectively. 
	In addition, we can observe that MF based methods (CMF, MC-BPR and BPR), achieve the worst performance on the Beibei dataset, which has more complicated and richer behaviors than the Tmall dataset. This  confirms the incapability of MF in modeling complicated interactions between users and items, being inferior to the multi-layer neural networks.
\vspace{0.1cm}
	\item \textbf{NMTR is a better framework than MC.}
	For each NCF model, we find that optimizing it under our NMTR framework outperforms optimizing it under the Multi-Channel BPR framework.
	Specifically, NMTR-NeuMF outperforms MC-NeuMF by 6.08\% on Beibei dataset and 30.76\% on Tmall dataset in HR@100. Thus, we can conclude that NMTR performs better than the MC framework in adapting a single-behavior recommender model for multiple behaviors. 
	
	To better understand the difference between two frameworks, we present the training loss and the testing performance in each training iteration in Figure~\ref{fig:overall1} (for Beibei) and Figure~\ref{fig:overall2} (for Tmall). In our NMTR framework, the training loss is defined as the joint loss in multi-task learning, which is a combination of the prediction loss of behaviors of multiple types. 
	We can observe that, for both datasets, although training loss of NMTR is the highest, it essentially demonstrates the best generalization performance. 
	For the Beibei dataset, we find that the HR score of MC-NeuMF starts to decrease after 40 iterations, even though the $L_2$ regularization and dropout have been adopted.
	Note that in Table~\ref{tab:Comparision}, we have reported the peak performance of each baseline evaluated per iteration (such a setting is to fully explore the potential of all methods). 
    Even so, our NMTR still outperforms MC-NeuMF by 6.08\% in HR@100 and 5.70\% in NDCG@100. 
    However, on the Tmall dataset, in which only two behaviors are available and the data is of a smaller scale, MC-NeuMF fails to utilize the view behavior to improve the performance (i.e., underperforms NeuMF). In contrast, our NMTR-NeuMF outperforms NeuMF by 30.76\% in HR@100 and 14.37\% in NDCG@100, which are very significant improvements. We also present the average training time per epoch of three models in Table~\ref{tab:trainingtime} and we can find our proposed NMTR framework's efficiency is acceptable.
	
\vspace{0.1cm}
	\item \textbf{The performance on multiple behaviors are relevant to that on single behavior.}
	No matter which framework is chosen, NMTR or MC, we can observe that the performance of the multi-behavior setting is relevant to that of single-behavior. 
	This is because that they use the same set of CF functions, which on the other hand implies that the performance on multi-behavior data maybe limited by the choice of the CF function. 
	An empirical evidence is that NMTR-MLP performs the worst among the three NMTR methods, which can be caused by the poor performance of MLP in modeling CF effect (in single-behavior data). 
	\textcolor{black}{In addition, for some metrics, such as HR@50 and NDCG@50 on both dataset, and HR@80 and NDCG@80 on Tmall dataset, NMTR-MLP and NMTR-GMF are outperformed by some baseline methods such as MC-NeuMF. It can be explained that MC-NeuMF's relatively better performance is due to NeuMF's best performance compared all single-behavior methods. Therefore, our NMTR-NeuMF achieves better performance than MC-NeuMF on these metrics.
	}
	Moreover, another important finding is that auxiliary behaviors could adversely degrade the performance without a proper modeling. An evidence can be found in the results of the Tmall dataset, where the methods under the MC framework fail to improve the performance in general. 
\end{itemize}
\vspace{-0.2cm}
To summarize, the extensive comparison on two real datasets verify that our proposed NMTR solution can effectively leverage multiple types of behaviors to improve the recommendation performance, i.e. our model outperforms the best baseline method by 6.08\% and 30.76\% on two datasets, respectively.

\begin{table*}[h]
	\begin{center}
	\small
		\caption{Performance of NMTR model with different combination of interaction data on the Beibei dataset}\label{tab:DataCombination}
		\begin{tabular}{|c|c|c|c|c|c|c|c|c|}
			\hline
			{}& \multicolumn{8}{|c|}{\bf Beibei Dataset}   \\\hline
			{\bf Interaction Subset}    & \multicolumn{2}{|c|}{\bf (Purchase, Carting)}& \multicolumn{2}{|c|}{\bf (Purchase, View)}& \multicolumn{2}{|c|}{\bf (Purchase, 50\% Carting) }  & \multicolumn{2}{|c|}{\bf (Purchase, 50\% View) }  \\\hline
			{\bf Performance }    & \bf HR@100  & \bf NDCG@100&  \bf HR@100  & \bf NDCG@100 &  \bf HR@100  & \bf NDCG@100 &  \bf HR@100  & \bf NDCG@100\\\hline

			\bf NMTR-GMF     &0.2979  &0.0705  &0.3029  &0.0726 &0.2947  &0.0701 &0.2953  &0.0698 \\\hline
			\bf NMTR-MLP &0.2770  &0.0670  &0.3140  &0.0741 &0.2726  &0.0654 &0.3058  &0.0725\\\hline
			\bf NMTR-NeuMF     &0.2882  &0.0691 &0.3147 &0.0743 &0.2778 &0.0676 &0.3107  &0.0737\\\hline

		\end{tabular}
		
	\end{center}
\end{table*}

\subsection{Impact of Auxiliary Behaviors and Parameters (RQ2)}

\textcolor{black}{
In order to understand how auxiliary behavior data affect the recommendation performance, we choose the Beibei dataset for further investigation since it has more types of behaviors. 
Since the motivation of multi-behavior recommendation is to utilize interaction data of other types of behaviors to help improving recommendation quality on target behavior, we
investigate how the data quality of auxiliary behaviors affects our NMTR model's performance.
An intuitive experimental setting is that to random sample auxiliary behaviors for our utilized two datasets while keeping target behaivor (i.e. purchase) intact.
Table~\ref{tab:DataCombination} shows the performance of different combinations of behavioral data. There are four sampling rules for obtaining a subset. For example, (Purchase, 50\%view) means that intact purchase records are kept and half records of view behavior are randomly selected to be kept for each user.
As mentioned above, when investigating top-$K$ performance, K=$100$ is a reasonable setting. Thus, here we evaluated the performance via two metrics: HR@100 and NDCG@100. We tuned hyper-parameters, with a similar way as Section~\ref{Sec:ExpSet}, to report the best performance for various subsets of interaction data.}
From the results, we have the following two observations.

First, adding views data leads to better performance than adding carts data. 
The main reason is probably that the cart data contains too similar signal with the purchase data and provides fewer new signal on user preference. 
Specifically, a purchase record is often accompanied by a carting record. 
On the contrary, the view behaviors provide some useful intermediate feedback such as, viewed and not bought, which can effectively improve the learning on binary implicit feedback.

Second, by using only 50\% of the cart and view interactions, we find that the performance is worse than the previous two experiments. 
Specifically, the performance of (Purchase, 50\% Carting) is worse than only using purchase, 
while (Purchase, 50\% View) is better than only using purchase. There are two major reasons. 
On one hand, view is the weakest signal to reflect user preference and the total number of views is very large, making the missing of part of view data is acceptable. 
Therefore, missing of some view records shall not affect the result too much. On the other hand, random missing of carts records can bring some noises, as cart behavior is very similar with the purchase behavior, and this validates the hypothesis in~\cite{steck2010training}: those missing records of some behaviors are more likely taken as negative value rather than missing value by model.

\begin{figure}[t]
	\centering
	\subfigure[NMTR-GMF on Beibei Dataset]{
		\label{fig:GMF_Coeff}
		\includegraphics[width=0.23\textwidth]{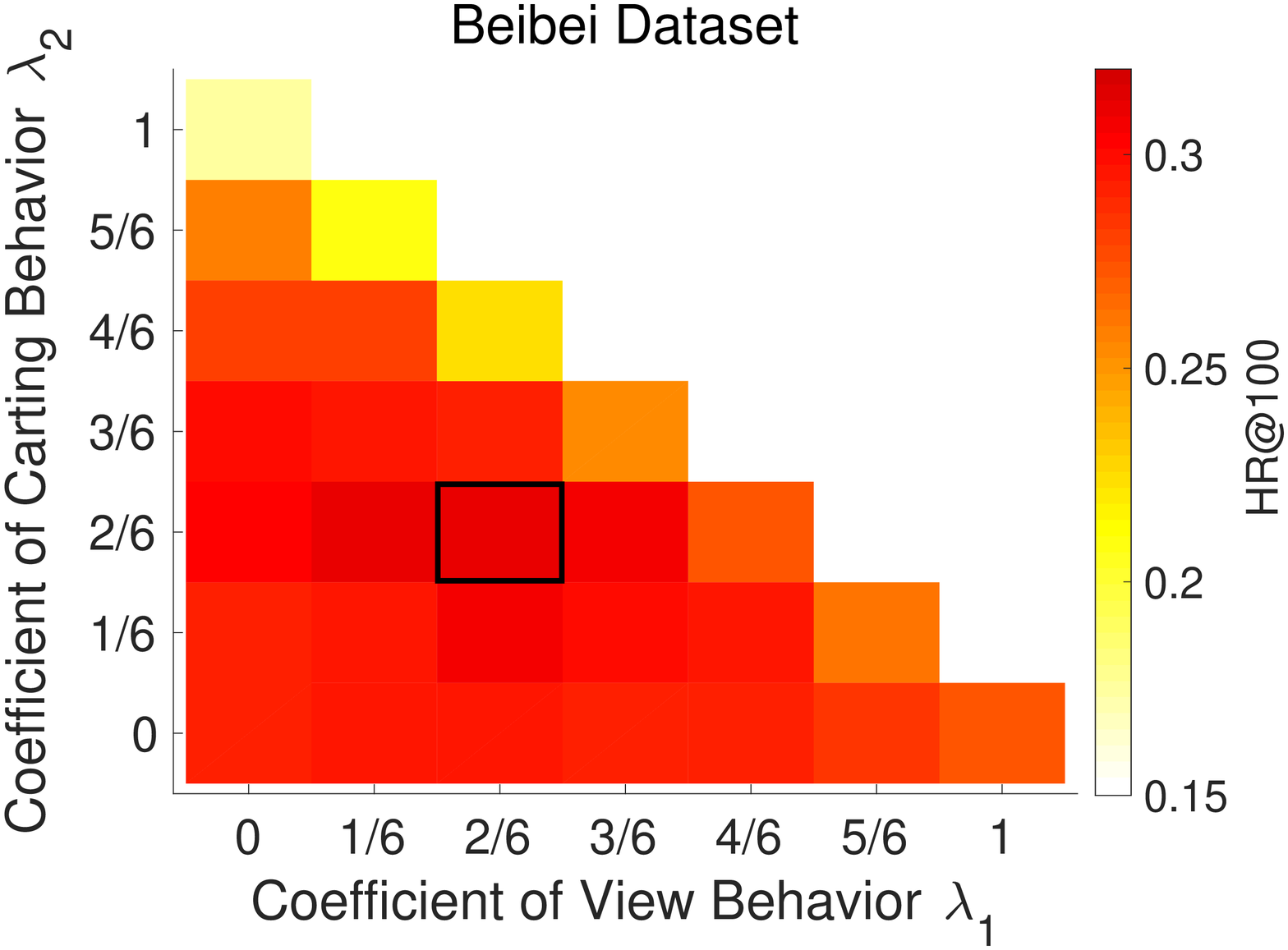}}
	\subfigure[NMTR-MLP on Beibei Dataset]{
		\label{fig:MLP_Coeff}
		\includegraphics[width=0.23\textwidth]{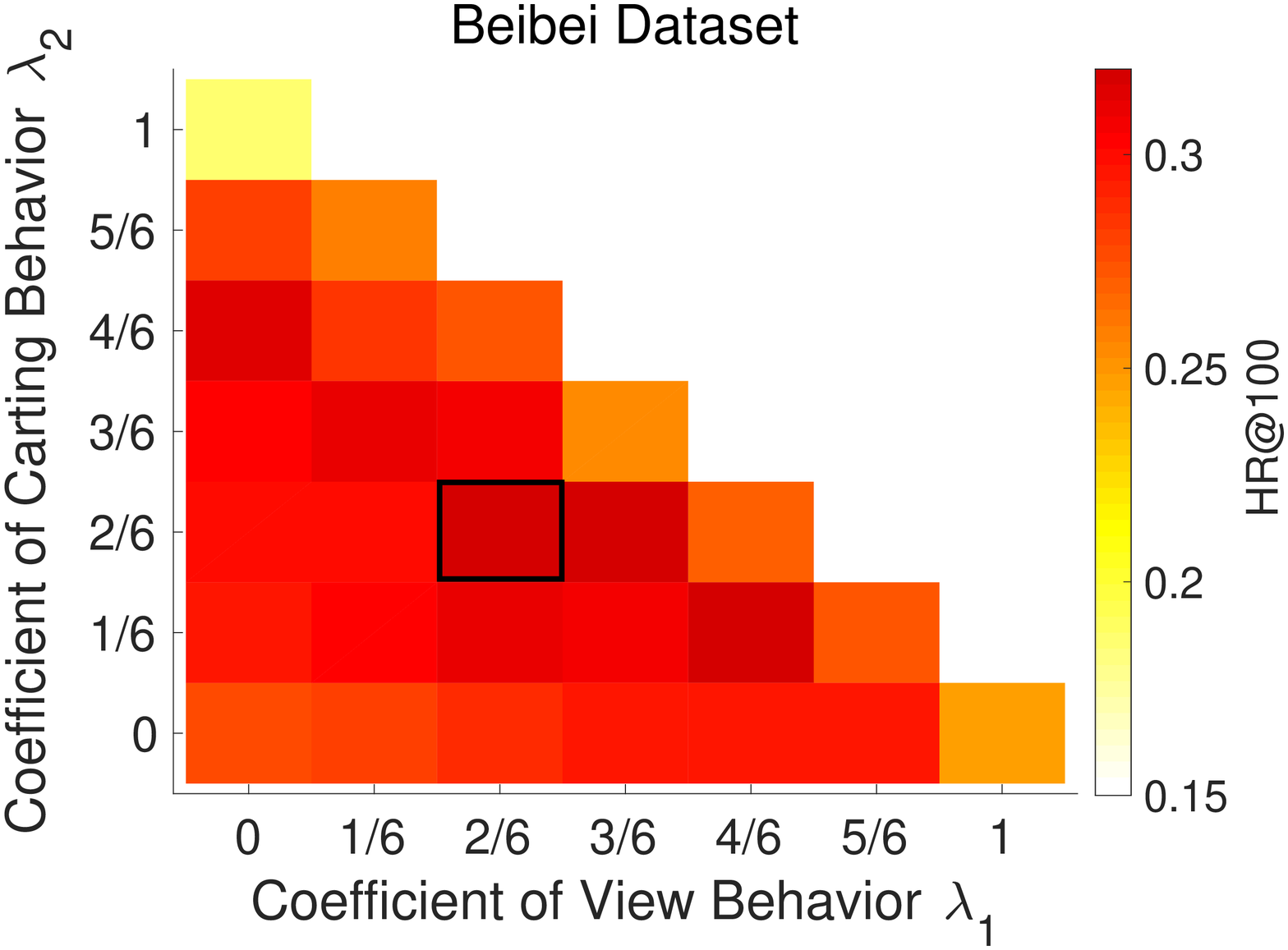}}	
	\subfigure[NMTR-NeuMF on Beibei Dataset]{
    	\label{fig:NeuMF_Coeff}
    	\includegraphics[width=0.23\textwidth]{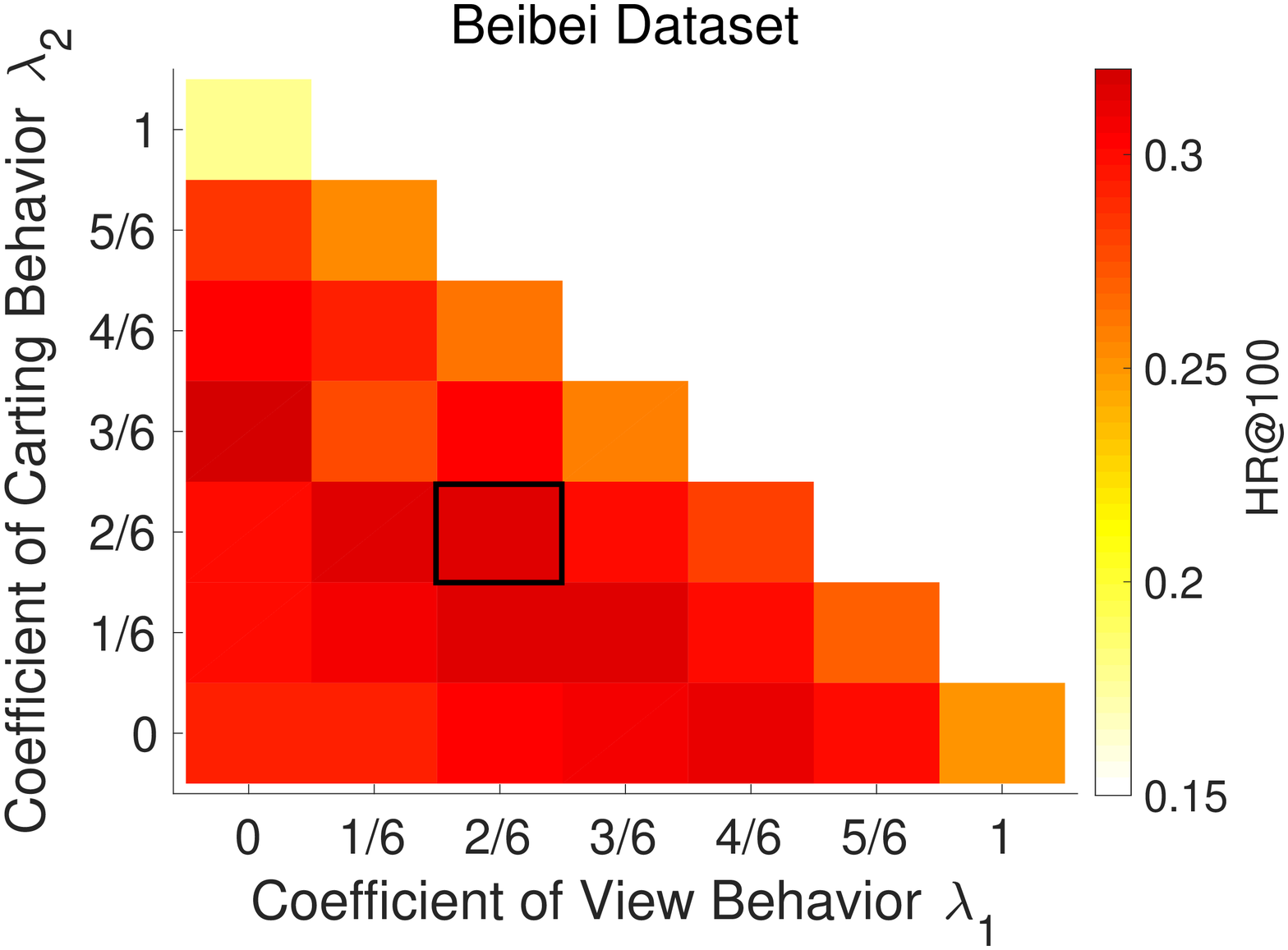}}
    \subfigure[NMTR on Beibei Dataset]{
    	\label{fig:Tmall_Coeff}
    	\includegraphics[width=0.23\textwidth]{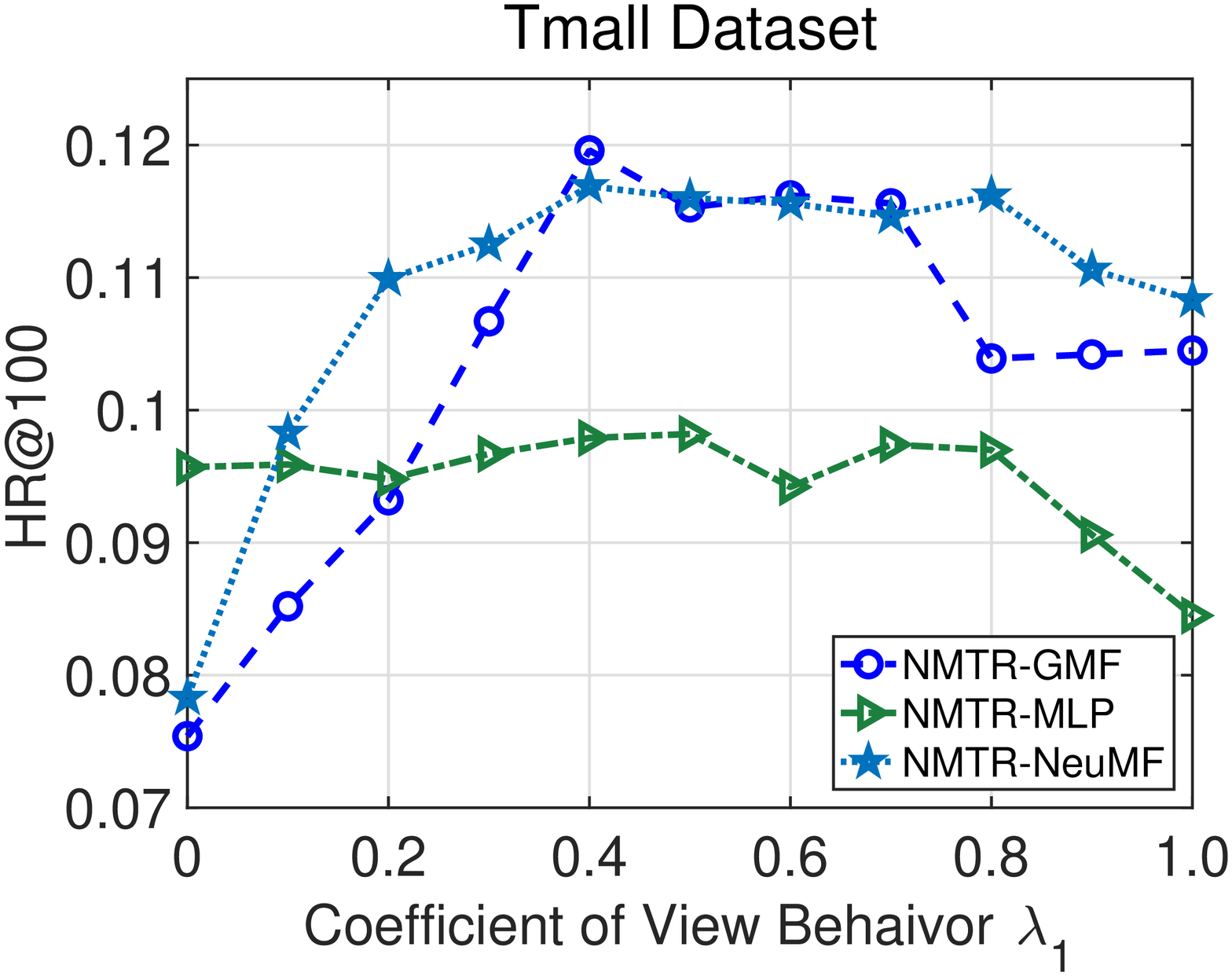}}	
	\caption{HR@100 Performance of NMTR with different loss coefficient on the Beibei and Tmall datasets}
	\label{Fig:Sparse}
\end{figure}

\textcolor{black}{
In order to understand how hyper-parameters impact the performance, we focus on the coefficient in the joint loss function of MTL, $\lambda_r$, since it controls the weight of each type of behavior and is a key parameter of our method.
There are three and two behavior types for Beibei and Tmall, respectively. 
For the Beibei dataset, there are three types of behaviors (view, cart and purchase), which means there are three loss coefficients $\lambda_{1}$, $\lambda_{2}$ and $\lambda_{3}$, respectively. Note that $\lambda_{1} + \lambda_{2} + \lambda_{3}=1$, we tune the three coefficients in $[0,\frac{1}{6},\frac{2}{6},\frac{3}{6},\frac{4}{6},\frac{5}{6},1]$ and plot the performance of HR@100 in Figure~\ref{fig:GMF_Coeff}, ~\ref{fig:MLP_Coeff} and~\ref{fig:NeuMF_Coeff}. When $\lambda_{1}$ and $\lambda_{2}$ are given, then value of $\lambda_{3}$ is determined. Therefore each block represents a setting of $\lambda_{r}$.
And in these three figures, darker blocks means better performance.
Similarly, for the Tmall dataset, there are only two types of behaviors (view and purchase), so there are two coefficients: $\lambda_{1} + \lambda_{2} = 1$. We tune $\lambda_{1}$ from $0$ to $1$ with step size $0.1$ and plot the HR@100 performance in Figure~\ref{fig:Tmall_Coeff}. }
For both datasets, the best performance of the NMTR methods are achieved at almost the same setting, (2/6,2/6,2/6) for the Beibei dataset and about(0.4, 0.6) for the Tmall dataset, which verifies that it is not so independent on the utilized CF unit. 
For Beibei dataset, in Figure~\ref{fig:GMF_Coeff}, ~\ref{fig:MLP_Coeff} and~\ref{fig:NeuMF_Coeff}, upper-right blocks are rather shallow since they represent a relatively small $\lambda_3$ which is the coefficient of purchase behavior. However, for Tmall dataset, in Figure~\ref{fig:Tmall_Coeff}, a relatively low coefficient of purchase behavior outperforms that of view behavior. We argue that it is due to size difference of auxiliary behavioral data in two datasets.

Furthermore, as mentioned in Section~\ref{Sec:MTL}, we utilize multi-task learning rather than sequential learning to optimize our proposed model.
Then to study how multi-task learning outperforms the intuitive sequential learning in optimizing the cascaded prediction models, we compare the performance of the two training methods in Table~\ref{tab:training}. Here we still adopt HR@100 and NDCG@100 as evaluation metrics. For the sequential learning, we feed the cascaded CF units for each behavior type with separated samples following the order of behaviors. We can find that training in the sequential manner achieved worse performance, which verified the effectiveness of our proposed multi-task training component.

\begin{table}[t]
    \begin{center}
        \scriptsize
              \begin{adjustwidth}{0.3cm}{}
        \caption{Performance comparison of sequential training and multi-task learning on the Beibei and Tmall datasets}\label{tab:training}
        \begin{tabular}{|c|c|c|c|c|}
            \hline
            {\bf Dataset} &\multicolumn{2}{|c|}{\bf Beibei}  &\multicolumn{2}{|c|}{\bf Tmall} \\\hline

            {\bf Performance} & \bf HR@100&\bf  NDCG@100 &\bf HR@100 &\bf NDCG@100 \\\hline         
            {\bf NMTR-GMF} &0.3119 &0.0741 &0.1196 &0.0314 \\\hline
            {\bf Sequential-GMF}&0.2730 &0.0672 &0.0913 &0.0290 \\\hline
            {\bf NMTR-MLP} &0.3188 &0.0762 &0.0982 &0.0290\\\hline
            {\bf Sequential-MLP} &0.2663 &0.0692 &0.0856 &0.0226\\\hline
            {\bf NMTR-NeuMF} &0.3193 &0.0760 &0.1169 &0.0366 \\\hline
{\bf Sequential-NeuMF} &0.2704 &0.0658 &0.0946 &0.0304\\\hline
\end{tabular}
\end{adjustwidth}
\end{center}
\end{table}

In summary, our NMTR incorporates the semantics of different behavior interactions and capture the ordinal relations among them. In addition, coefficient $\lambda_r$, as a significant hyper-parameter in our NMTR model, is independent with CF unit. Furthermore, multi-task training is demonstrated to far better than sequential training.

\subsection{Impact of Data Sparsity (RQ3)}

Data sparsity is a big challenge for recommender systems based on implicit feedbacks~\cite{kabbur2013fism, yin2017mobi}, and multi-behavior recommendation is a typical solution of it. Thus, we study how our proposed NMTR model improves the recommendation for those users having few records of target behavior. Specifically, we divided all users of the Beibei dataset into several groups according to the number of purchase records: [5-8, 9-12, 13-16, 17-20, $>$20]. In each group, the number of users are in the range of 4000 to 5000, which 
eliminates the randomness of experimental results.
For each group, we compare the performance of our methods with baseline methods. For NMTR and MC models, we only plot the most competitive ones, NMTR-NeuMF and MC-NeuMF, for clarity; for baselines for single-behavior data, we also only plot the best one, NeuMF. 

\begin{figure}[t]
	\centering
	\subfigure[HR@100]{
		\label{fig:RQ4_HR}
		\includegraphics[width=0.23\textwidth]{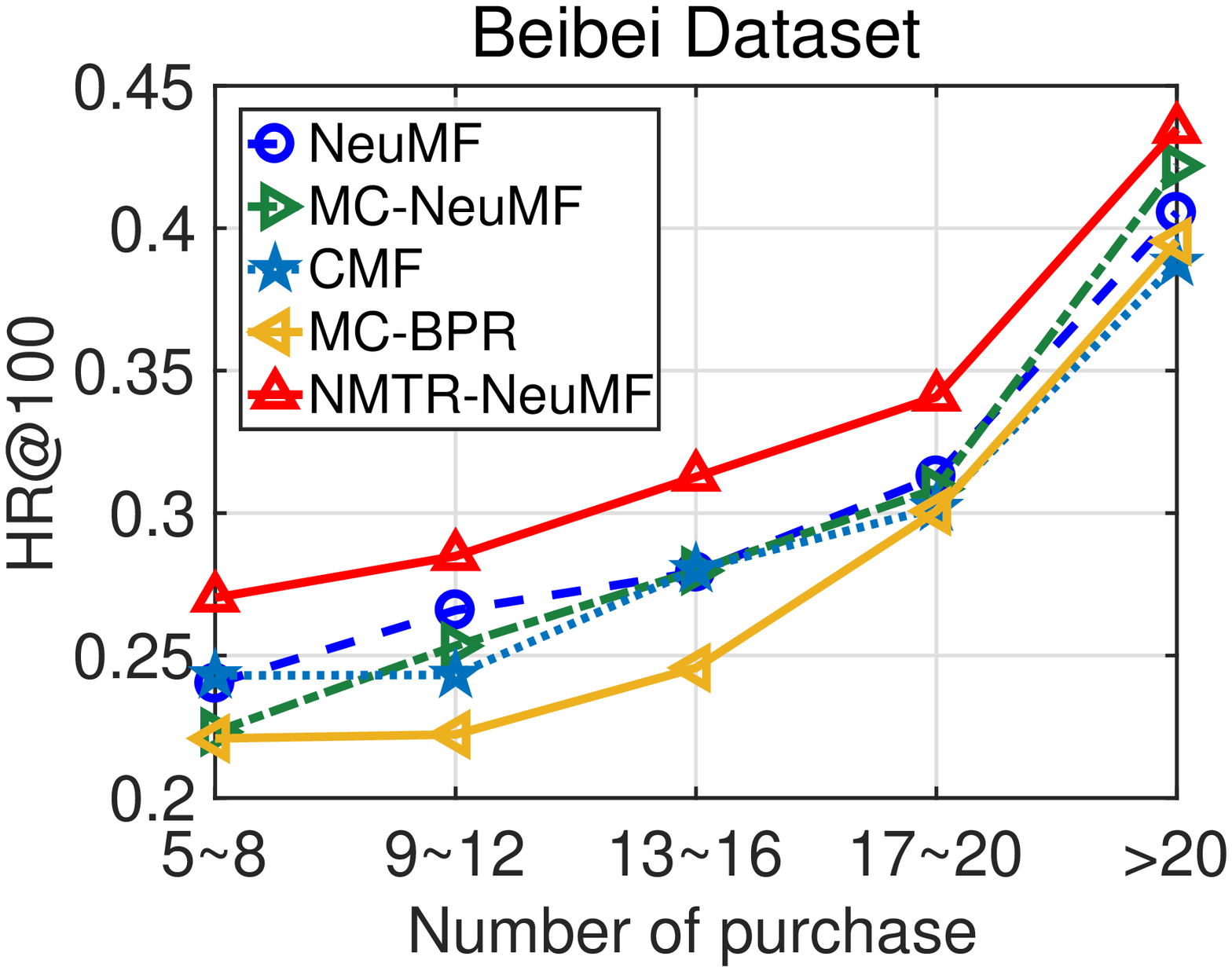}}
	\subfigure[NDCG@100]{
		\label{fig:RQ4_NDCG}
		\includegraphics[width=0.23\textwidth]{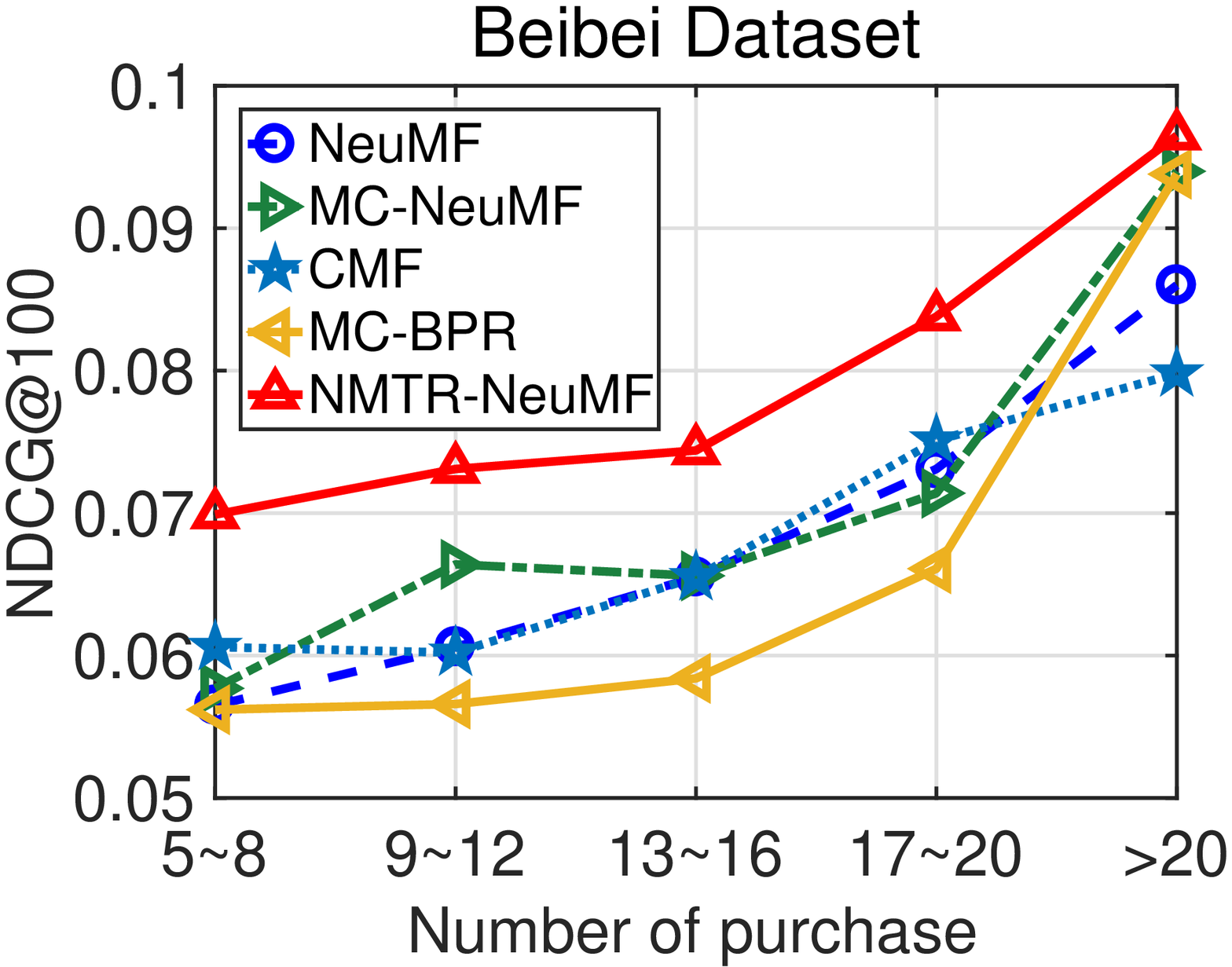}}	
	\caption{Performance of NeuMF, MC-NeuMF, CMF, MC-BPR and NMTR-NeuMF on users with different number of purchase records}
	\label{Fig:Sparse}
\end{figure}
The results are shown in Figure~\ref{Fig:Sparse}.
From the results, we can observe that when the user purchase data becomes sparser, the recommendation performance of NMTR-NeuMF decreases slower than other methods. 
Especially for NDCG, from fifth to first user group, NMTR-NeuMF is decreased by 27.56\% while MC-BPR and MC-NeuMF is decreased by 40.09\% and 38.62\%.
Furthermore, even in the first user group with only 5-8 purchase records, our NMTR still keeps a good recommendation performance of 0.027 for HR@100 and 0.07 for NDCG@100, which outperforms the best baseline by 11.23\% and 15.35\%, respectively. As a result, the performance gap between NMTR and other methods becomes larger when data become sparser. Since NMTR model learns the other type of behaviors in a reasonable way, it can achieve a good performance for users with sparse interactions. As a summary, we conclude that our proposed NMTR model solves data sparsity problem efficiently to some extent.

In conclusion, we conduct extensive experiments on two real-word datasets, which verifies that our proposed NMTR model outperform existing recommendation methods. Further studies demonstrate our model can alleviate data sparisty problem efficiently.

%% file: 5.relatedwork.tex
\section{Related Work}\label{sec:related}

\subsection{{Multi-Behavior Recommendation.}}

In today's online information systems, user can interact with a system in multiple forms. There are many works~\cite{lo2016understanding, dong2019brand} analyzing and modeling such multiple types of behaviors. Lo~\textit{et al.}~\cite{lo2016understanding} studied the influence of click and save behavior on the final purchase decision via case study. Moe~\textit{et al.}~\cite{moe2004dynamic}, Dong~\textit{et al.}~\cite{dong2019brand} and Lee~\textit{et al.}~\cite{lee2015online} utilized various time-evolving behavioral features to predict purchase behaviors. Olbrich~\textit{et al.}~\cite{olbrich2011modeling} and Yehezki~\textit{et al.}~\cite{yehezki2017classifying} proposed to extract features from user clickstreams to help predict purchase. These works verify the effectiveness of other types of behaviors to help model the target behavior.
Multi-behavior based recommendation aims to leverage the behavior data of other types to improve the recommendation performance on the target behavior. Matrix factorization, a prevalent method for single-behavior based recommendation~\cite{rendle2009bpr,kabbur2013fism}, has been adapted to the multi-behavior scenario. Ajit \textit{et al.}~\cite{CMF} first proposed a collective matrix factorization model (CMF) to simultaneously factorize multiple user-item interaction matrices with sharing item-side embeddings across matrices.
 Some other works extended the CMF to handle datasets of multiple user behaviors~\cite{zhao2015improving,krohn2012multi}. 
 Zhe \textit{et al.}~\cite{zhao2015improving} considered different behaviors in online social network (comment, re-share, and create-post), while Artus \textit{et al.}~\cite{krohn2012multi} extended CMF with sharing user-side embeddings in recommendation based social network data.
 On the other hand, some works approach multi-behavior recommendation from the perspective of learning~\cite{loni2016bayesian,qiu2018bprh,guo2017resolving,ding2018improving}. Babak \textit{et al.}~\cite{loni2016bayesian} proposed an extension of Bayesian Personalized Ranking (BPR)~\cite{rendle2009bpr}, as Multi-channel BPR, to adapt the sampling rule from different types of behavior in training of standard BPR. 
 Qiu~\text{et al.}~\cite{qiu2018bprh} proposed an adaptive sampler for BPR considering co-occurrence of multiple feedbacks while Guo~\textit{et al.}~\cite{guo2017resolving} proposed to sample unobserved items as positive items based on item-item similarity, which is calculated by multiple types of feedbacks. Ding~\textit{et al.}~\cite{ding2018improving} developed a margin-based pairwise learning framework when view-data is available.
 As discussed in the introduction, these existing models suffer from several limitations, which are addressed by our neural network-based solution NMTR.

\vspace{-0.30cm}
\subsection{{Neural Network Based Recommendation.}} Salakhutdinov \textit{et al.}~\cite{salakhutdinov2007restricted} proposed a Restricted Boltzmann Machines to predict explicit ratings, which was the first to apply neural network to recommender system. Recently, lots of works utilize neural network to extract the auxiliary information and features in recommender system, such as textual~\cite{zheng2017joint, tang2017joint,cheng2018aspect,cheng2019mmalfm,cheng20183ncf}, visual~\cite{mcauley2015image,wang2017your}, audio~\cite{van2013deep, van2013deep} and video~\cite{covington2016deep}. Rather than these other side features, some other works make use of recurrent neural network to model temporal features in recommender system~\cite{wu2016personal,okura2017embedding, song2016multi,kang2018self}. 

More recently, He \textit{et al.} \cite{he2017neural} proposed a neural network architecture for collaborative filtering, named Neural Collaborative Filtering (NCF), which learns the user-item interaction function using neural networks. It has been extended to adapt to different recommendation scenarios~\cite{wang2017item, chen2017attentive}. For example, Wang \textit{et al.} \cite{wang2017item} applied NCF to model user-item interaction in both information domain and social domain, and Chen \textit{et al.} \cite{chen2017attentive} combined NCF with attention mechanism to recommend videos and images. Recently, inspire the advances in graph representation learning, some works~\cite{ying2018graph,GCNSocialRec,wang2019neural} utilize graph neural network~\cite{kipf2016semi} for recommendation tasks.
 Our work extends the architecture of NCF to a multi-task learning framework, which aims to solve the problem of learning recommender systems from multi-behavior data.

\subsection{{Multi-task Learning for Recommendation.}} 
In multi-task learning (MTL) framework, various related tasks can share common representations, while training in parallel.  Traditional multi-task learning works are mainly based on matrix regularization~\cite{evgeniou2004regularized, argyriou2007multi} and neural-based approach~\cite{jiang2018exploiting,yang2016deep}. 
To the best of our knowledge,~\cite{ning2010multi} is the first work to apply multi-task learning to recommender system, which built a MTL framework to limit the similarity between users and similarity between items. Bansal \textit{et al.}~\cite{bansal2016ask} proposed a gated-recurrent-units based MTL network which share the embedded representation of texts and output personalized text for different users. In contrast, our work adapts MTL our task to effectively learn from multiple user behaviors.

%% file: 0.head.bbl
\begin{thebibliography}{10}
\providecommand{\url}[1]{#1}
\csname url@samestyle\endcsname
\providecommand{\newblock}{\relax}
\providecommand{\bibinfo}[2]{#2}
\providecommand{\BIBentrySTDinterwordspacing}{\spaceskip=0pt\relax}
\providecommand{\BIBentryALTinterwordstretchfactor}{4}
\providecommand{\BIBentryALTinterwordspacing}{\spaceskip=\fontdimen2\font plus
\BIBentryALTinterwordstretchfactor\fontdimen3\font minus
  \fontdimen4\font\relax}
\providecommand{\BIBforeignlanguage}[2]{{%
\expandafter\ifx\csname l@#1\endcsname\relax
\typeout{** WARNING: IEEEtran.bst: No hyphenation pattern has been}%
\typeout{** loaded for the language `#1'. Using the pattern for}%
\typeout{** the default language instead.}%
\else
\language=\csname l@#1\endcsname
\fi
#2}}
\providecommand{\BIBdecl}{\relax}
\BIBdecl

\bibitem{rendle2009bpr}
S.~Rendle, C.~Freudenthaler, Z.~Gantner, and L.~Schmidt-Thieme, ``Bpr: Bayesian
  personalized ranking from implicit feedback,'' in \emph{UAI}, 2009, pp.
  452--461.

\bibitem{he2017neural}
X.~He, L.~Liao, H.~Zhang, L.~Nie, X.~Hu, and T.-S. Chua, ``Neural collaborative
  filtering,'' in \emph{WWW}, 2017, pp. 173--182.

\bibitem{CMF}
A.~P. Singh and G.~J. Gordon, ``Relational learning via collective matrix
  factorization,'' in \emph{SIGKDD}.\hskip 1em plus 0.5em minus 0.4em\relax
  ACM, 2008, pp. 650--658.

\bibitem{park2017also}
C.~Park, D.~Kim, J.~Oh, and H.~Yu, ``Do also-viewed products help user rating
  prediction?'' in \emph{WWW}, 2017, pp. 1113--1122.

\bibitem{krohn2012multi}
A.~Krohn-Grimberghe, L.~Drumond, C.~Freudenthaler, and L.~Schmidt-Thieme,
  ``Multi-relational matrix factorization using bayesian personalized ranking
  for social network data,'' in \emph{WSDM}, 2012, pp. 173--182.

\bibitem{zhao2015improving}
Z.~Zhao, Z.~Cheng, L.~Hong, and E.~H. Chi, ``Improving user topic interest
  profiles by behavior factorization,'' in \emph{WWW}, 2015, pp. 1406--1416.

\bibitem{loni2016bayesian}
B.~Loni, R.~Pagano, M.~Larson, and A.~Hanjalic, ``Bayesian personalized ranking
  with multi-channel user feedback,'' in \emph{RecSys}, 2016, pp. 361--364.

\bibitem{qiu2018bprh}
H.~Qiu, Y.~Liu, G.~Guo, Z.~Sun, J.~Zhang, and H.~T. Nguyen, ``Bprh: Bayesian
  personalized ranking for heterogeneous implicit feedback,'' \emph{Information
  Sciences}, vol. 453, pp. 80--98, 2018.

\bibitem{Zhang:2014:EFM}
Y.~Zhang, G.~Lai, M.~Zhang, Y.~Zhang, Y.~Liu, and S.~Ma, ``Explicit factor
  models for explainable recommendation based on phrase-level sentiment
  analysis,'' in \emph{SIGIR}, 2014, pp. 83--92.

\bibitem{heckel2017scalable}
R.~Heckel, M.~Vlachos, T.~Parnell, and C.~D{\"u}nner, ``Scalable and
  interpretable product recommendations via overlapping co-clustering,'' in
  \emph{ICDE}, 2017, pp. 1033--1044.

\bibitem{kabbur2013fism}
S.~Kabbur, X.~Ning, and G.~Karypis, ``Fism: factored item similarity models for
  top-n recommender systems,'' in \emph{SIGKDD}, 2013, pp. 659--667.

\bibitem{wang2017item}
X.~Wang, X.~He, L.~Nie, and T.-S. Chua, ``Item silk road: Recommending items
  from information domains to social users,'' in \emph{SIGIR}, 2017.

\bibitem{RL_review}
Y.~Bengio, A.~Courville, and P.~Vincent, ``Representation learning: A review
  and new perspectives,'' \emph{TPAMI}, vol.~35, no.~8, pp. 1798--1828, 2013.

\bibitem{NNCF}
T.~Bai, J.~Wen, J.~Zhang, and W.~X. Zhao, ``A neural collaborative filtering
  model with interaction-based neighborhood,'' in \emph{CIKM}, 2017, pp.
  1979--1982.

\bibitem{xiang2010temporal}
L.~Xiang, Q.~Yuan, S.~Zhao, L.~Chen, X.~Zhang, Q.~Yang, and J.~Sun, ``Temporal
  recommendation on graphs via long-and short-term preference fusion,'' in
  \emph{Proceedings of the 16th ACM SIGKDD international conference on
  Knowledge discovery and data mining}.\hskip 1em plus 0.5em minus 0.4em\relax
  ACM, 2010, pp. 723--732.

\bibitem{rendle2010factorizing}
S.~Rendle, C.~Freudenthaler, and L.~Schmidt-Thieme, ``Factorizing personalized
  markov chains for next-basket recommendation,'' in \emph{Proceedings of the
  19th international conference on World wide web}.\hskip 1em plus 0.5em minus
  0.4em\relax ACM, 2010, pp. 811--820.

\bibitem{hidasi2015session}
B.~Hidasi, A.~Karatzoglou, L.~Baltrunas, and D.~Tikk, ``Session-based
  recommendations with recurrent neural networks,'' in \emph{ICLR}, 2016.

\bibitem{evgeniou2004regularized}
T.~Evgeniou and M.~Pontil, ``Regularized multi--task learning,'' in
  \emph{SIGKDD}, 2004, pp. 109--117.

\bibitem{DBLP:conf/sigir/ChenZAXYQ17}
X.~Chen, Y.~Zhang, Q.~Ai, H.~Xu, J.~Yan, and Z.~Qin, ``Personalized key frame
  recommendation,'' in \emph{SIGIR}, 2017, pp. 315--324.

\bibitem{he2015delving}
K.~He, X.~Zhang, S.~Ren, and J.~Sun, ``Delving deep into rectifiers: Surpassing
  human-level performance on imagenet classification,'' in \emph{ICCV}, 2015,
  pp. 1026--1034.

\bibitem{kingma2014adam}
D.~Kingma and J.~Ba, ``Adam: A method for stochastic optimization,'' in
  \emph{ICLR}, 2015.

\bibitem{duchi2011adaptive}
J.~Duchi, E.~Hazan, and Y.~Singer, ``Adaptive subgradient methods for online
  learning and stochastic optimization,'' \emph{JMLR}, vol.~12, no. Jul, pp.
  2121--2159, 2011.

\bibitem{wang2016spore}
W.~Wang, H.~Yin, S.~Sadiq, L.~Chen, M.~Xie, and X.~Zhou, ``Spore: A sequential
  personalized spatial item recommender system,'' in \emph{ICDE}, 2016, pp.
  954--965.

\bibitem{steck2010training}
H.~Steck, ``Training and testing of recommender systems on data missing not at
  random,'' in \emph{SIGKDD}, 2010, pp. 713--722.

\bibitem{yin2017mobi}
H.~Yin, L.~Chen, W.~Wang, X.~Du, Q.~V.~H. Nguyen, and X.~Zhou, ``Mobi-sage: A
  sparse additive generative model for mobile app recommendation,'' in
  \emph{ICDE}, 2017, pp. 75--78.

\bibitem{lo2016understanding}
C.~Lo, D.~Frankowski, and J.~Leskovec, ``Understanding behaviors that lead to
  purchasing: A case study of pinterest,'' in \emph{Proceedings of the 22nd ACM
  SIGKDD international conference on knowledge discovery and data
  mining}.\hskip 1em plus 0.5em minus 0.4em\relax ACM, 2016, pp. 531--540.

\bibitem{dong2019brand}
Y.~Dong and W.~Jiang, ``Brand purchase prediction based on time-evolving user
  behaviors in e-commerce,'' \emph{Concurrency and Computation: Practice and
  Experience}, vol.~31, no.~1, p. e4882, 2019.

\bibitem{moe2004dynamic}
W.~W. Moe and P.~S. Fader, ``Dynamic conversion behavior at e-commerce sites,''
  \emph{Management Science}, vol.~50, no.~3, pp. 326--335, 2004.

\bibitem{lee2015online}
M.~Lee, T.~Ha, J.~Han, J.-Y. Rha, and T.~T. Kwon, ``Online footsteps to
  purchase: Exploring consumer behaviors on online shopping sites,'' in
  \emph{Proceedings of the ACM Web Science Conference}.\hskip 1em plus 0.5em
  minus 0.4em\relax ACM, 2015, p.~15.

\bibitem{olbrich2011modeling}
R.~Olbrich and C.~Holsing, ``Modeling consumer purchasing behavior in social
  shopping communities with clickstream data,'' \emph{International Journal of
  Electronic Commerce}, vol.~16, no.~2, pp. 15--40, 2011.

\bibitem{yehezki2017classifying}
S.~Yehezki and A.~Dhini, ``Classifying purchase decision based on user
  clickstream in e-commerce using web usage mining,'' in \emph{Proceedings of
  the International Conference on Business and Information Management}.\hskip
  1em plus 0.5em minus 0.4em\relax ACM, 2017, pp. 57--61.

\bibitem{guo2017resolving}
G.~Guo, H.~Qiu, Z.~Tan, Y.~Liu, J.~Ma, and X.~Wang, ``Resolving data sparsity
  by multi-type auxiliary implicit feedback for recommender systems,''
  \emph{Knowledge-Based Systems}, vol. 138, pp. 202--207, 2017.

\bibitem{ding2018improving}
J.~Ding, G.~Yu, X.~He, Y.~Quan, Y.~Li, T.-S. Chua, D.~Jin, and J.~Yu,
  ``Improving implicit recommender systems with view data.'' in \emph{IJCAI},
  2018, pp. 3343--3349.

\bibitem{salakhutdinov2007restricted}
R.~Salakhutdinov, A.~Mnih, and G.~Hinton, ``Restricted boltzmann machines for
  collaborative filtering,'' in \emph{ICML}, 2007, pp. 791--798.

\bibitem{zheng2017joint}
L.~Zheng, V.~Noroozi, and P.~S. Yu, ``Joint deep modeling of users and items
  using reviews for recommendation,'' in \emph{WSDM}, 2017, pp. 425--434.

\bibitem{tang2017joint}
L.~Tang and E.~Y. Liu, ``Joint user-entity representation learning for event
  recommendation in social network,'' in \emph{ICDE}.\hskip 1em plus 0.5em
  minus 0.4em\relax IEEE, 2017, pp. 271--280.

\bibitem{cheng2018aspect}
Z.~Cheng, Y.~Ding, L.~Zhu, and M.~Kankanhalli, ``Aspect-aware latent factor
  model: Rating prediction with ratings and reviews,'' in \emph{Proceedings of
  the 2018 World Wide Web Conference}.\hskip 1em plus 0.5em minus 0.4em\relax
  International World Wide Web Conferences Steering Committee, 2018, pp.
  639--648.

\bibitem{cheng2019mmalfm}
Z.~Cheng, X.~Chang, L.~Zhu, R.~C. Kanjirathinkal, and M.~Kankanhalli, ``Mmalfm:
  Explainable recommendation by leveraging reviews and images,'' \emph{ACM
  Transactions on Information Systems (TOIS)}, vol.~37, no.~2, p.~16, 2019.

\bibitem{cheng20183ncf}
Z.~Cheng, Y.~Ding, X.~He, L.~Zhu, X.~Song, and M.~S. Kankanhalli, ``A\^{} 3ncf:
  An adaptive aspect attention model for rating prediction.'' in \emph{IJCAI},
  2018, pp. 3748--3754.

\bibitem{mcauley2015image}
J.~McAuley, C.~Targett, Q.~Shi, and A.~Van Den~Hengel, ``Image-based
  recommendations on styles and substitutes,'' in \emph{SIGIR}, 2015, pp.
  43--52.

\bibitem{wang2017your}
S.~Wang, Y.~Wang, J.~Tang, K.~Shu, S.~Ranganath, and H.~Liu, ``What your images
  reveal: Exploiting visual contents for point-of-interest recommendation,'' in
  \emph{WWW}, 2017, pp. 391--400.

\bibitem{van2013deep}
A.~Van~den Oord, S.~Dieleman, and B.~Schrauwen, ``Deep content-based music
  recommendation,'' in \emph{NIPS}, 2013, pp. 2643--2651.

\bibitem{covington2016deep}
P.~Covington, J.~Adams, and E.~Sargin, ``Deep neural networks for youtube
  recommendations,'' in \emph{RecSys}, 2016, pp. 191--198.

\bibitem{wu2016personal}
S.~Wu, W.~Ren, C.~Yu, G.~Chen, D.~Zhang, and J.~Zhu, ``Personal recommendation
  using deep recurrent neural networks in netease,'' in \emph{ICDE}, 2016, pp.
  1218--1229.

\bibitem{okura2017embedding}
S.~Okura, Y.~Tagami, S.~Ono, and A.~Tajima, ``Embedding-based news
  recommendation for millions of users,'' in \emph{SIGKDD}, 2017, pp.
  1933--1942.

\bibitem{song2016multi}
Y.~Song, A.~M. Elkahky, and X.~He, ``Multi-rate deep learning for temporal
  recommendation,'' in \emph{SIGIR}, 2016, pp. 909--912.

\bibitem{kang2018self}
W.-C. Kang and J.~McAuley, ``Self-attentive sequential recommendation,'' in
  \emph{2018 IEEE International Conference on Data Mining (ICDM)}.\hskip 1em
  plus 0.5em minus 0.4em\relax IEEE, 2018, pp. 197--206.

\bibitem{chen2017attentive}
J.~Chen, H.~Zhang, X.~He, L.~Nie, W.~Liu, and T.-S. Chua, ``Attentive
  collaborative filtering: Multimedia recommendation with item-and
  component-level attention,'' in \emph{SIGIR}, 2017, pp. 335--344.

\bibitem{ying2018graph}
R.~Ying, R.~He, K.~Chen, P.~Eksombatchai, W.~L. Hamilton, and J.~Leskovec,
  ``Graph convolutional neural networks for web-scale recommender systems,'' in
  \emph{Proceedings of the 24th ACM SIGKDD International Conference on
  Knowledge Discovery \& Data Mining (KDD)}.\hskip 1em plus 0.5em minus
  0.4em\relax ACM, 2018, pp. 974--983.

\bibitem{GCNSocialRec}
W.~Fan, Y.~Ma, Q.~Li, Y.~He, E.~Zhao, J.~Tang, and D.~Yin, ``Graph neural
  networks for social recommendation,'' in \emph{The World Wide Web Conference
  (WWW)}, 2019, pp. 417--426.

\bibitem{wang2019neural}
X.~Wang, X.~He, M.~Wang, F.~Feng, and T.-S. Chua, ``Neural graph collaborative
  filtering,'' in \emph{International Conference on Research and Development in
  Information Retrieval (SIGIR)}, 2019.

\bibitem{kipf2016semi}
T.~N. Kipf and M.~Welling, ``Semi-supervised classification with graph
  convolutional networks,'' in \emph{International Conference on Learning
  Representations (ICLR)}, 2017.

\bibitem{argyriou2007multi}
A.~Argyriou, T.~Evgeniou, and M.~Pontil, ``Multi-task feature learning,'' in
  \emph{NIPS}, 2007, pp. 41--48.

\bibitem{jiang2018exploiting}
Y.-G. Jiang, Z.~Wu, J.~Wang, X.~Xue, and S.-F. Chang, ``Exploiting feature and
  class relationships in video categorization with regularized deep neural
  networks,'' \emph{TPAMI}, vol.~40, no.~2, pp. 352--364, 2018.

\bibitem{yang2016deep}
Y.~Yang and T.~Hospedales, ``Deep multi-task representation learning: A tensor
  factorisation approach,'' in \emph{ICLR}, 2017.

\bibitem{ning2010multi}
X.~Ning and G.~Karypis, ``Multi-task learning for recommender system,'' in
  \emph{ACML}, 2010, pp. 269--284.

\bibitem{bansal2016ask}
T.~Bansal, D.~Belanger, and A.~McCallum, ``Ask the gru: Multi-task learning for
  deep text recommendations,'' in \emph{RecSys}, 2016, pp. 107--114.

\end{thebibliography}
